\documentclass[conference,10pt]{IEEEtran}
\usepackage{booktabs}
\usepackage{amsmath}
\usepackage{dcolumn}
\usepackage{graphicx}
\usepackage{amsfonts}
\usepackage{url}
\usepackage{csquotes}
\usepackage{flushend}

\newcolumntype{d}[1]{D{.}{.}{#1}}
\newcommand{\softmax}{\text{softmax}}
\newcommand{\loss}{\text{loss}}

\begin{document}
\title{Towards Evaluating the Robustness \\ of Neural Networks}
\author{Nicholas Carlini \qquad David Wagner  \\ University of California, Berkeley}
\date{}
\maketitle

\section*{Abstract}

Neural networks provide state-of-the-art results for most machine learning tasks.
Unfortunately, neural networks are vulnerable to adversarial examples:
given an input $x$ and any target classification $t$, it is possible to find
a new input $x'$ that is similar to $x$ but classified as $t$. This makes it
difficult to apply neural networks in security-critical areas.
Defensive distillation is a recently proposed approach that
can take an arbitrary neural network, and increase its robustness,
reducing the success rate of current attacks' ability to find adversarial examples from $95\%$ to
$0.5\%$.

In this paper, we demonstrate that defensive distillation does not significantly
increase the robustness of
neural networks by introducing three new attack algorithms that are
successful on both distilled and undistilled neural networks with $100\%$ probability.
Our attacks are tailored to three distance metrics used previously in the literature,
and when compared to previous adversarial example generation algorithms,
our attacks are often much more effective (and never worse). Furthermore, we propose
using high-confidence adversarial examples in a simple transferability test
we show can also be used to break defensive distillation.
We hope our attacks will be used as a benchmark in future defense attempts
to create neural networks that resist adversarial examples.

\section{Introduction}

Deep neural networks have become increasingly effective at
many difficult machine-learning tasks. In the image recognition domain,
they are able to recognize images with near-human accuracy \cite{lecun1998gradient,
krizhevsky2012imagenet}.
They are also used for speech recognition \cite{hinton2012deep},
natural language processing \cite{andor2016globally}, and playing games
\cite{silver2016mastering, mnih2013playing}.

However, researchers have discovered that existing neural networks
are vulnerable to attack.
Szegedy \emph{et al.} \cite{szegedy2013intriguing} 
first noticed the existence of \emph{adversarial examples} in the image classification
domain: it is possible to transform an image by a small amount
and thereby change how the image is classified.
Often, the total amount of change required can be so small as to be undetectable.

The degree to which attackers can find adversarial examples limits the domains
in which neural networks can be used. For example, if we use neural networks in
self-driving cars, adversarial examples could allow an attacker to cause the
car to take unwanted actions.

The existence of adversarial examples
has inspired research on how to harden neural networks
against these kinds of attacks. Many early attempts to secure neural networks
failed or provided only marginal robustness improvements \cite{gu2014towards,bastani2016measuring,
huang2015learning,shaham2015understanding}.

\begin{figure}
  \textbf{\,\,\,Original\,\,\, Adversarial \,\,\,\,\,\,\,\,\,\, Original\,\,\, Adversarial}\par\medskip
  \centering
  \includegraphics[scale=0.125]{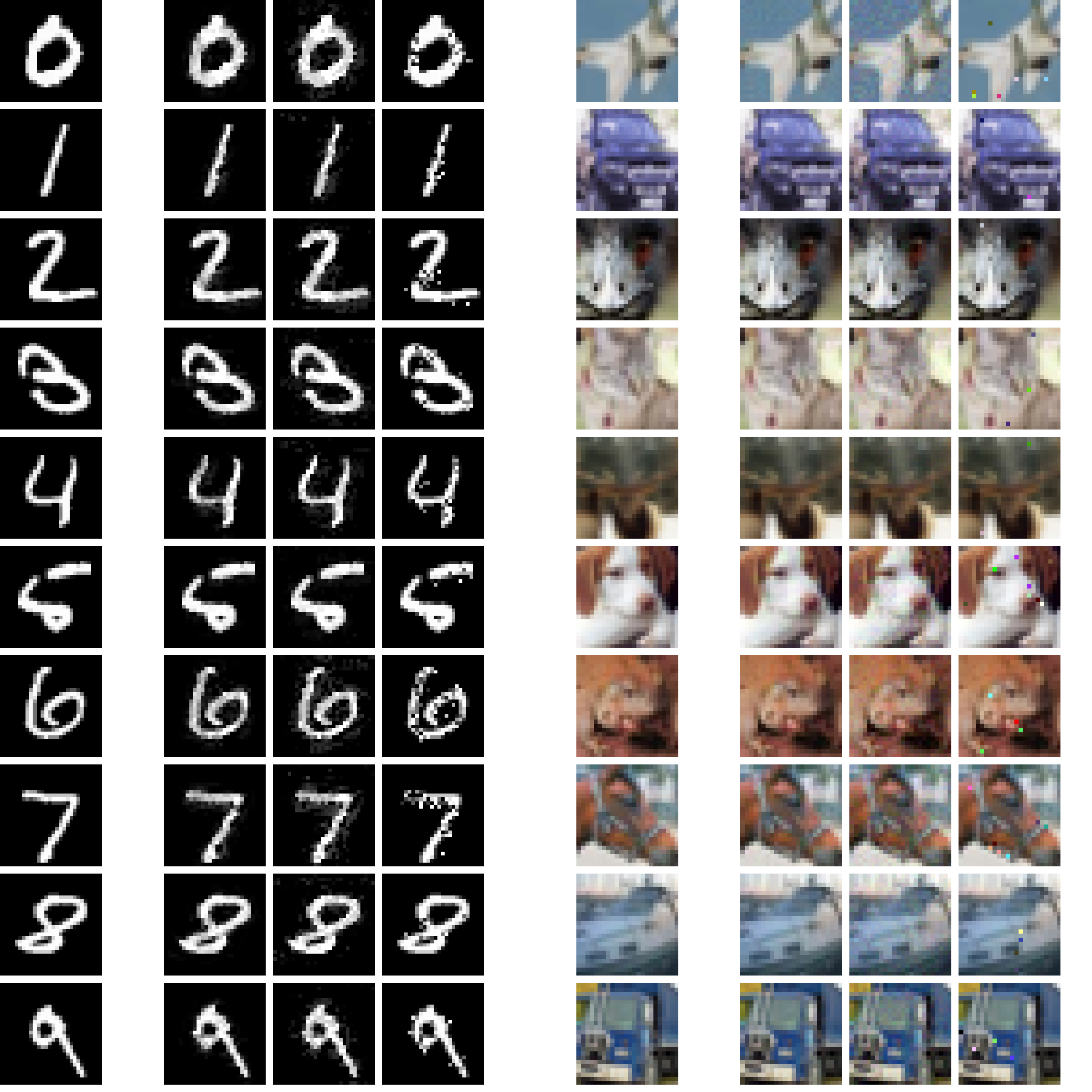}
  \caption{An illustration of our attacks on a defensively distilled network.  The leftmost column contains
  the starting image.  The next three columns
  show adversarial examples generated by our $L_2$, $L_{\infty}$,
  and $L_0$ algorithms, respectively.
  All images start out classified correctly with label $l$, and the three misclassified instances
  share the same misclassified label of $l+1 \pmod{10}$. Images were chosen as the
  first of their class from the test set.}
  \label{fig:example-attacks}
\end{figure}

\emph{Defensive distillation} \cite{distillation} is one such recent
defense proposed for hardening neural networks against adversarial
examples.
Initial analysis proved to be very promising: defensive distillation defeats existing
attack algorithms and reduces their success probability
from $95\%$ to $0.5\%$. 
Defensive distillation can be applied to any feed-forward neural network and only
requires a single re-training step, and is currently one of the only defenses
giving strong security guarantees against adversarial examples.

In general, there are two different approaches one can take to evaluate the robustness of a neural
network: attempt to prove a lower bound, or construct attacks that
demonstrate an upper bound. The former approach, while sound, is substantially more
difficult to implement in practice, and all attempts have required approximations
\cite{bastani2016measuring,huang2016safety}.
On the other hand, if the attacks used in the the latter approach are not
sufficiently strong and fail often, the upper bound may not be useful.

In this paper we create a set of attacks that can be used to construct an upper bound
on the robustness of neural networks. As a case study, we use these attacks to
demonstrate that defensive distillation does not actually
eliminate adversarial examples. We construct three new attacks (under three
previously used distance metrics: $L_0$, $L_2$, and $L_\infty$) that
succeed in finding adversarial examples for $100\%$ of images on defensively
distilled networks. While defensive distillation stops previously published
attacks, it cannot resist the more powerful attack techniques we introduce
in this paper.

This case study illustrates the general need for better techniques to evaluate
the robustness of neural networks: while distillation was shown to be secure
against the current state-of-the-art attacks, it fails against our stronger attacks.
Furthermore, when comparing our attacks against the current
state-of-the-art on standard unsecured models, our methods generate adversarial
examples with less total distortion in every case.
We suggest that our attacks are a better baseline for evaluating
candidate defenses: before placing any faith in a new possible defense,
we suggest that designers at least check whether it can resist our attacks.

We additionally propose using high-confidence adversarial examples to evaluate
the robustness of defenses. Transferability
\cite{szegedy2013intriguing,goodfellow2014explaining} is the well-known property that
adversarial examples on one model are often also adversarial on another model.
We demonstrate that adversarial examples from our attacks are transferable from
the unsecured model
to the defensively distilled (secured) model. In general, we argue that any
defense must demonstrate it is able to break the transferability property.

We evaluate our attacks on three standard datasets: MNIST \cite{lecun1998mnist}, a
digit-recognition task (0-9); CIFAR-10 \cite{krizhevsky2009learning}, a small-image
recognition task, also with 10 classes; and ImageNet \cite{deng2009imagenet}, a
large-image recognition task with 1000 classes.

Figure~\ref{fig:example-attacks} shows examples of adversarial examples our
techniques generate on defensively distilled networks trained on the
MNIST and CIFAR datasets.

In one extreme example for the ImageNet classification
task, we can cause the Inception v3 \cite{szegedy2015rethinking}
network to incorrectly classify images by changing only the lowest order
bit of each pixel. Such changes are impossible to detect visually.

To enable others to more easily use our work to evaluate the robustness
of other defenses, all of our adversarial example generation algorithms (along with
code to train the models we use, to reproduce the results we present) are
available online at \url{http://nicholas.carlini.com/code/nn_robust_attacks}.

This paper makes the following contributions:
\begin{itemize}
\item We introduce three new attacks for the $L_0$, $L_2$,
  and $L_\infty$ distance metrics.
  Our attacks are significantly more effective than
  previous approaches. Our $L_0$ attack is the first published
  attack that can cause targeted misclassification
  on the ImageNet dataset.
\item We apply these attacks to defensive distillation and discover
  that distillation provides little security benefit over un-distilled networks.
\item We propose using high-confidence adversarial examples
  in a simple transferability test to evaluate defenses, and show this
  test breaks defensive distillation.
\item We systematically evaluate the choice of the objective function for finding
  adversarial examples, and show that the choice can dramatically impact the
  efficacy of an attack.
\end{itemize}

\section{Background}

\subsection{Threat Model}

Machine learning is being used in an increasing array of settings to make
potentially security critical decisions: self-driving cars
\cite{bojarski2016end, selfdriving},
drones \cite{giusti2016machine},
robots \cite{mnih2015human,janglova2005neural},
anomaly detection \cite{chandola2009anomaly},
malware classification \cite{dahl2013large,pascanu2015malware,yuan2014droid},
speech recognition and recognition of voice commands \cite{38131,graves2013speech},
NLP \cite{andor2016globally},
and many more. Consequently, understanding the security properties of deep learning
has become a crucial question in this area. The extent to which we can construct
adversarial examples influences the settings in which we may want to (or not want to)
use neural networks.

In the speech recognition domain, recent work has shown \cite{carlini2016hidden} it is possible to
generate audio that sounds like speech to machine learning algorithms but not to humans.
This can
be used to control user's devices without their knowledge. For example, by playing
a video with a hidden voice command, it may be possible to cause a smart phone to
visit a malicious webpage to cause a drive-by download. This work focused on conventional
techniques (Gaussian Mixture Models and Hidden Markov Models), but as speech recognition
is increasingly using neural networks, the study of adversarial examples becomes
relevant in this domain. \footnote{Strictly speaking, hidden voice commands are not
  adversarial examples because they are not similar to the original input \cite{carlini2016hidden}.}

In the space of malware classification, the existence of adversarial examples
not only limits their potential application settings, but entirely defeats its
purpose: an adversary who is able to make only slight modifications to a malware
file that cause it to remain malware, but become classified as benign, has
entirely defeated the malware classifier \cite{dahl2013large,grosse2016adversarial}.

Turning back to the threat to self-driving cars introduced earlier,
this is not an
unrealistic attack: it has been shown that
adversarial examples are possible in the physical world \cite{kurakin2016adversarial}
after taking pictures of them.

The key question then becomes exactly how much distortion we must add to
cause the classification to change. In each domain, the distance metric
that we must use is different. In the space of images, which we focus on in this
paper, we rely on previous work that suggests that various $L_p$ norms are reasonable
approximations of human perceptual distance (see Section~\ref{sec:metric} for more information).

We assume in this paper that the adversary has complete access to a neural network, including
the architecture and all paramaters, and can use this in a white-box
manner. This is a conservative and realistic assumption: prior work has shown it
is possible to train a substitute model given black-box access to a target model,
and by attacking the substitute model, we can then transfer these attacks to the
target model. \cite{papernot2016transferability}

Given these threats, there have been various attempts
\cite{gu2014towards,bastani2016measuring,
  huang2015learning,shaham2015understanding,distillation}
at constructing defenses that increase
the \emph{robustness} of a neural network, defined as a measure of how easy it is
to find adversarial examples that are close to their original input.

In this paper we study one of these, \emph{distillation as a defense} \cite{distillation},
that hopes to secure an
arbitrary neural network.
This type of defensive distillation was shown to make generating adversarial examples
nearly impossible for existing attack techniques \cite{distillation}.
We find that although the current state-of-the-art fails to find adversarial
examples for defensively distilled networks, the stronger attacks we develop in this
paper \emph{are} able to construct adversarial examples.

\subsection{Neural Networks and Notation}

A neural network is a function $F(x) = y$ that accepts an input $x \in \mathbb{R}^n$
and produces an output $y \in \mathbb{R}^m$.
The model $F$ also implicitly depends on some model parameters $\theta$; in our work
the model is fixed, so for convenience we don't show the dependence on $\theta$.

In this paper we focus on neural networks used as an $m$-class classifier.
The output of the network is computed using the softmax function,
which ensures that the output vector $y$ satisfies
$0 \le y_i \le 1$ and $y_1 + \dots + y_m = 1$.
The output vector $y$ is thus treated as a probability distribution, i.e.,
$y_i$ is treated as the probability that input $x$ has class $i$.
The classifier assigns the label $C(x) = \arg\max_i F(x)_i$ to the input $x$.
Let $C^*(x)$ be the correct label of $x$.
The inputs to the softmax function are called \emph{logits}.

We use the notation from Papernot et al. \cite{distillation}: define $F$ to 
be the full neural network including the softmax function, $Z(x) = z$ to be the output of
all layers except the softmax (so $z$ are the logits), and
\begin{equation*}
F(x) = \softmax(Z(x)) = y.
\end{equation*}
A neural network typically \footnote{Most simple networks have this simple
  linear structure, however other more sophisticated networks have
  more complicated structures (e.g., ResNet \cite{he2016deep} and Inception \cite{szegedy2015rethinking}).
  The network architecture does not impact our attacks.}
consists of layers
\begin{equation*}
F = \softmax \circ F_n \circ F_{n-1} \circ \cdots \circ F_1
\end{equation*}
where
\begin{equation*}
F_i(x)  = \sigma(\theta_i \cdot x) + \hat\theta_i
\end{equation*}
for some non-linear activation function $\sigma$, some matrix $\theta_i$ of model
weights, and some vector $\hat\theta_i$ of model biases. Together $\theta$ and
$\hat\theta$ make up the model parameters.
Common choices of $\sigma$ 
are tanh \cite{mishkin2015all}, sigmoid, ReLU \cite{maas2013rectifier}, or ELU \cite{clevert2015fast}. 
In this paper we focus primarily on networks that use a ReLU activation function,
as it currently is the most widely used
activation function
\cite{szegedy2015rethinking,springenberg2014striving,mishkin2015all,distillation}.

We use image classification as our primary evaluation domain.
An $h\times w$-pixel grey-scale image is a two-dimensional vector $x \in \mathbb{R}^{hw}$,
where $x_i$ denotes the intensity of pixel $i$ and is scaled to be in the range $[0,1]$.
A color RGB image is a three-dimensional vector $x \in \mathbb{R}^{3hw}$.
We do
not convert RGB images to HSV, HSL, or other cylindrical coordinate representations of 
color images: the neural networks act on raw pixel values.

\subsection{Adversarial Examples}
Szegedy \emph{et al.} \cite{szegedy2013intriguing} first pointed out the
existence of \emph{adversarial examples}: given a valid input $x$ and a target
$t \ne C^*(x)$, it is often
possible to find a similar input $x'$ such that $C(x') = t$ yet
$x,x'$ are close according to some distance metric. An example $x'$ with
this property is known as a \emph{targeted} adversarial example.

A less powerful attack also discussed in the literature instead asks for
\emph{untargeted} adversarial examples: instead of classifying $x$ as a given target
class, we only search for an input $x'$
so that $C(x') \ne C^*(x)$ and $x,x'$ are close.
Untargeted attacks are strictly less powerful than targeted attacks and we do not
consider them in this paper. \footnote{An untargeted attack is simply a
  more efficient (and often less accurate) method of running a
  targeted attack for each target and taking the closest. In this paper we focus
  on identifying the most accurate attacks, and do not consider untargeted attacks.}


Instead, we consider three different approaches for
how to choose the target class, in a targeted attack:
\begin{itemize}
\item \emph{Average Case:} select the target class \emph{uniformly at random} among the
  labels that are not the correct label.
\item \emph{Best Case:} perform the attack against all incorrect classes, and report
  the target class that was \emph{least difficult} to attack.
\item \emph{Worst Case:} perform the attack against all incorrect classes, and report
  the target class that was \emph{most difficult} to attack.
\end{itemize}
In all of our evaluations we perform all three types of attacks: best-case,
average-case, and worst-case. Notice that if a classifier is only accurate
$80\%$ of the time, then the best case attack will require a change of
$0$ in $20\%$ of cases.

On ImageNet, we approximate the best-case and worst-case attack by sampling 100
random target classes out of the 1,000 possible for efficiency reasons.

\subsection{Distance Metrics}
\label{sec:metric}
In our definition of adversarial examples, we require use of a distance metric
to quantify similarity.
There are three widely-used distance metrics in the
literature for generating adversarial examples, all of which
are $L_p$ norms.


The $L_p$ distance is written $\|x-x'\|_p$, where
the $p$-norm $\|\cdot\|_p$ is defined as
\begin{equation*}
\|v\|_p = \left(\sum\limits_{i=1}^n \lvert v_i \rvert ^p\right)^{1\over p}.
\end{equation*}
In more detail:
\begin{enumerate}
\item $L_0$ distance measures the number of coordinates $i$ such that
$x_i \ne x'_i$.
Thus, the $L_0$ distance corresponds to the number of pixels that
have been altered in an image.\footnote{In RGB images, there are
three channels that each can change. We count the number of \emph{pixels} that
are different, where two pixels are considered different if
\emph{any} of the three colors are different. We do not consider a distance metric 
where an attacker can change one color plane but not another meaningful. We relax this
requirement when comparing to other $L_0$ attacks that do not make this assumption to
provide for a fair comparison.}

Papernot \emph{et al.} argue for the use of the $L_0$ distance metric, and it is the primary
distance metric under which defensive distillation's security is argued \cite{distillation}.

\item $L_2$ distance measures the standard Euclidean (root-mean-square) distance
between $x$ and $x'$. The $L_2$ distance can remain small when there are many small
changes to many pixels.

This distance metric was used in the initial adversarial example work \cite{szegedy2013intriguing}.

\item $L_{\infty}$ distance measures the maximum
change to any of the coordinates:
\begin{equation*}
\|x-x'\|_\infty = \max (|x_1-x'_1|,\dots,|x_n-x'_n|).
\end{equation*}
For images, we can imagine there is a maximum budget, and each pixel is
allowed to be changed by up to this limit, with no limit on the number of
pixels that are modified.

Goodfellow \emph{et al.} argue that $L_\infty$ is the optimal distance metric to use
\cite{warde2016adversarial} and in a follow-up paper Papernot \emph{et al.} argue distillation is
secure under this distance metric \cite{papernot2016effectiveness}.

\end{enumerate}

No distance metric is a perfect measure of human perceptual similarity,
and we pass no judgement on exactly which distance metric is optimal. We believe
constructing and evaluating a good distance metric is an important research question
we leave to future work.

However, since most existing work has picked one of these three distance metrics,
and since defensive distillation argued security against two of these, we too use these
distance metrics and construct attacks that perform superior to the state-of-the-art
for each of these distance metrics.

When reporting all numbers in this paper, we report using the distance
metric as defined above, on the range $[0,1]$. (That is, changing a pixel
in a greyscale image
from full-on to full-off will result in $L_2$ change of $1.0$ and a
$L_\infty$ change of $1.0$, not $255$.)

\subsection{Defensive Distillation}

We briefly provide a high-level overview of defensive distillation. We provide a
complete description later in Section~\ref{sec:distillation}.

To defensively distill a neural network, begin by first training a network
with identical architecture on the training data in a standard manner.
When we compute the softmax while training this network,  replace it with a
more-smooth version of the softmax (by dividing the logits by some constant $T$).
At the end of training, generate the \emph{soft training labels} by evaluating this
network on each of the training instances and taking the output labels of the
network.

Then, throw out the first network and use only the soft training labels.
With those, train a second network where instead of training it on the original
training labels, use the soft labels. This trains the second model to behave
like the first model, and the soft labels convey additional hidden knowledge
learned by the first model.

The key insight here is that by training to match the first network, we will hopefully
avoid over-fitting against any of the training data. If the reason that neural networks
exist is because neural networks are highly non-linear and have ``blind spots''
\cite{szegedy2013intriguing}
where adversarial examples lie, then preventing this type of over-fitting might remove
those blind spots.

In fact, as we will see later, defensive distillation does not remove adversarial
examples. One potential reason this may occur is that others
\cite{goodfellow2014explaining} have argued the
reason adversarial examples exist is not due to blind spots in a highly non-linear
neural network, but due only to the locally-linear nature of neural networks. This
so-called linearity hypothesis appears to be true \cite{warde2016adversarial}, and under this
explanation it is perhaps less surprising that distillation does not increase the
robustness of neural networks.

\subsection{Organization}
The remainder of this paper is structured as follows.
In the next section, we survey existing attacks that have been proposed in the
literature for generating adversarial examples, for the $L_2$, $L_\infty$, and $L_0$
distance metrics.
We then describe our attack algorithms that target the same three distance metrics and
provide superior results to the prior work.
Having developed these attacks, we 
review defensive distillation in more detail and discuss why the existing
attacks fail to find adversarial examples on defensively distilled networks.
Finally, we attack defensive distillation with our new algorithms and show that
it provides only limited value.

\section{Attack Algorithms}

\subsection{L-BFGS}
Szegedy \emph{et al.} \cite{szegedy2013intriguing} generated adversarial examples
using box-constrained L-BFGS.
Given an image $x$, their method finds a different image $x'$ that is
similar to $x$ under $L_2$ distance, yet is labeled differently by the classifier.
They model the problem as a constrained minimization problem:
\begin{align*}
  \text{minimize } \;& \|x-x'\|_2^2\\
  \text{such that } \;& C(x') = l \\
    & x' \in [0,1]^n
\end{align*}
This problem can be very difficult to solve, however, so Szegedy \emph{et al.}
instead solve the following problem:
\begin{align*}
  \text{minimize } \;& c \cdot \|x-x'\|_2^2 + \loss_{F,l}(x')\\
  \text{such that } \;& x' \in [0,1]^n
\end{align*}
where $\loss_{F,l}$ is a function mapping an image to a positive real number.
One common loss function to use is cross-entropy.
Line search is performed to find the constant $c>0$ that yields an adversarial
example of minimum distance: in other words, we repeatedly solve this optimization
problem for multiple values of $c$, adaptively updating $c$
using bisection search or any other
method for one-dimensional optimization.

\subsection{Fast Gradient Sign}
The fast gradient sign \cite{goodfellow2014explaining} method has two key differences from the
L-BFGS method: first, it is optimized for the $L_{\infty}$ distance metric,
and second, it is designed primarily to be fast instead of producing very
close adversarial examples.
Given an image $x$ the fast gradient sign method sets
\begin{equation*}
  x' = x - \epsilon \cdot \text{sign}(\nabla \loss_{F,t}(x)),
\end{equation*}
where $\epsilon$ is chosen to be sufficiently small so as to be undetectable,
and $t$ is the target label.
Intuitively, for each pixel, the fast gradient sign method uses the gradient
of the loss function to determine in which direction
the pixel's intensity should be changed (whether it should be increased or
decreased) to minimize the loss function; then, it shifts all pixels
simultaneously.

It is important to note that the fast gradient sign attack
was designed to be \emph{fast}, rather than optimal. It is not meant to produce the
minimal adversarial perturbations.

\paragraph*{Iterative Gradient Sign}

Kurakin \emph{et al.} introduce a simple refinement of the fast gradient sign method
\cite{kurakin2016adversarial}
where instead of taking a single step of size $\epsilon$ in the direction of the
gradient-sign, multiple smaller steps $\alpha$ are taken, and the result is clipped by
the same $\epsilon$. Specifically, begin by setting
\begin{equation*}
  x'_0 = 0
\end{equation*}
and then on each iteration
\begin{equation*}
  x'_i = x'_{i-1} - \text{clip}_{\epsilon}(\alpha \cdot \text{sign}(\nabla \loss_{F,t}(x'_{i-1})))
\end{equation*}
Iterative gradient sign was found to produce superior results to fast gradient sign
\cite{kurakin2016adversarial}.

\subsection{JSMA}
Papernot \emph{et al.} introduced an attack optimized under $L_0$
distance \cite{papernot2016limitations} known as the
Jacobian-based Saliency Map Attack (JSMA).
We give a brief summary of their attack algorithm;
for a complete description and motivation, we encourage
the reader to read their original paper \cite{papernot2016limitations}.

At a high level, the attack
is a greedy algorithm that picks pixels to modify one at a
time, increasing the target classification on each iteration.
They use the gradient $\nabla Z(x)_l$ to compute a \emph{saliency map}, which
models the impact each pixel has on the resulting classification.
A large value indicates that
changing it will significantly increase the likelihood of the model
labeling the image as the target class $l$. Given the saliency map, it picks the
most important pixel and modify it to increase the likelihood of class $l$.
This is repeated until either more than a set threshold of pixels are modified
which makes the attack detectable, or it succeeds in changing the classification.

In more detail, we begin by defining the saliency map in terms of a pair of pixels
$p,q$. Define
\begin{align*}
  \alpha_{pq} & = \sum\limits_{i \in \{p,q\}} \frac{\partial Z(x)_t}{\partial x_i} \\
  \beta_{pq} & = \left(\sum\limits_{i \in \{p,q\}} \sum\limits_{j} \frac{\partial Z(x)_j}{\partial x_i}\right)-\alpha_{pq}
\end{align*}
so that $\alpha_{pq}$ represents how much changing both pixels $p$ and $q$ will change the target
classification, and $\beta_{pq}$ represents how much changing $p$ and $q$ will change
all other outputs. Then the algorithm picks
\begin{align*}
  (p^*, q^*) & = \arg\max\limits_{(p, q)}\  (-\alpha_{pq} \cdot \beta_{pq}) \cdot (\alpha_{pq} > 0) \cdot (\beta_{pq} < 0)
\end{align*}
so that $\alpha_{pq}>0$ (the target class
is more likely), $\beta_{pq}<0$ (the other classes become less likely),
and $-\alpha_{pq} \cdot \beta_{pq}$ is largest.

Notice that JSMA uses the output of the second-to-last
layer $Z$, the logits, in the calculation of the gradient: the
output of the softmax $F$ is \emph{not} used. We refer to this as the \textbf{JSMA-Z} attack.

However, when the authors apply this attack to their defensively distilled networks,
they modify the attack so it uses $F$ instead of $Z$.
In other words, their computation uses the output of the softmax ($F$)
instead of the logits ($Z$).
We refer to this modification as the \textbf{JSMA-F} attack.\footnote{We verified this via personal communication with the authors.}

When an image has multiple color channels (e.g., RGB), this attack considers
the $L_0$ difference to be $1$ for each color channel changed independently (so that
if all three color channels of one pixel change change, the $L_0$ norm would be 3). While we
do not believe this is a meaningful threat model, when comparing to this attack, we
evaluate under both models.


\subsection{Deepfool}
Deepfool \cite{moosavi2015deepfool} is an untargeted attack technique optimized for the $L_2$
distance metric. It is efficient and produces closer adversarial examples than
the L-BFGS approach discussed earlier.

The authors construct Deepfool by imagining that the neural networks are totally linear,
with a hyperplane separating each class from another. From this, they analytically
derive the optimal solution to this simplified problem, and construct the adversarial
example.

Then, since neural networks are not actually linear, they take a step towards that
solution, and repeat the process a second time. The search terminates when a true
adversarial example is found.

The exact formulation used is rather sophisticated; interested readers should refer to
the original work \cite{moosavi2015deepfool}.

\begin{table}[t]
\centering
\begin{tabular}{lll}
  \toprule
  Layer Type & MNIST Model & CIFAR Model \\
  \midrule
  Convolution + ReLU & 3$\times$3$\times$32 & 3$\times$3$\times$64 \\
  Convolution + ReLU & 3$\times$3$\times$32 & 3$\times$3$\times$64 \\
  Max Pooling & 2$\times$2 & 2$\times$2 \\
  Convolution + ReLU & 3$\times$3$\times$64 & 3$\times$3$\times$128 \\
  Convolution + ReLU & 3$\times$3$\times$64 & 3$\times$3$\times$128 \\
  Max Pooling & 2$\times$2 & 2$\times$2 \\
  Fully Connected + ReLU & 200 & 256 \\
  Fully Connected + ReLU & 200 & 256 \\
  Softmax & 10 & 10 \\
 \bottomrule
\end{tabular}
\vskip 0.1in
\caption{\textbf{Model architectures} for the MNIST and CIFAR models. This architecture is
identical to that of the original defensive distillation work. \cite{distillation}}
  \label{tbl:models}
\end{table}

\begin{table}[t]
\centering
\begin{tabular}{lll}
  \toprule
  Parameter & MNIST Model & CIFAR Model \\
  \midrule
  Learning Rate & 0.1 & 0.01 (decay 0.5) \\
  Momentum & 0.9 & 0.9 (decay 0.5) \\
  Delay Rate & - & 10 epochs \\
  Dropout & 0.5 & 0.5 \\
  Batch Size & 128 & 128 \\
  Epochs & 50 & 50 \\
 \bottomrule
\end{tabular}
\vskip 0.1in
\caption{\textbf{Model parameters} for the MNIST and CIFAR models. These parameters are
identical to that of the original defensive distillation work. \cite{distillation}}
  \label{tbl:modelsparams}
\end{table}

\section{Experimental Setup}
\label{sec:model}
Before we develop our attack algorithms to break distillation, we describe how we
train the models on which we will evaluate our attacks.

We train two networks for the MNIST \cite{lecun1998mnist} and CIFAR-10
\cite{krizhevsky2009learning} classification tasks,
and use one pre-trained network for the ImageNet classification task \cite{imagenet}.
Our models and training approaches are identical to those presented in \cite{distillation}.
We achieve $99.5\%$ accuracy on MNIST, comparable to the state of the art.
On CIFAR-10, we achieve $80\%$ accuracy, identical to the accuracy given
in the distillation work. \footnote{This is compared to the state-of-the-art
result of $95\%$ \cite{graham2014fractional,springenberg2014striving,mishkin2015all}.
However, in order to provide the most accurate comparison to the original work,
we feel it is important to reproduce their model architectures.}

\vspace{3mm}
\emph{MNIST and CIFAR-10.}
The model architecture is given in Table~\ref{tbl:models} and
the hyperparameters selected in Table~\ref{tbl:modelsparams}. We use a
momentum-based SGD optimizer during training.

The CIFAR-10 model significantly overfits the training
data even with dropout: we obtain a final training cross-entropy loss of $0.05$
with accuracy $98\%$, compared to a validation loss of $1.2$ with validation
accuracy $80\%$. We do not alter the network by performing image augmentation
or adding additional dropout as that was not done in \cite{distillation}.

\vspace{3mm}
\emph{ImageNet.}
Along with considering MNIST and CIFAR, which are both relatively small datasets,
we also consider the ImageNet dataset.
Instead of training our own ImageNet model, we use
the pre-trained Inception v3 network \cite{szegedy2015rethinking},
which achieves $96\%$ top-5 accuracy (that is, the probability that the
correct class is one of the five most likely as reported by the network
is $96\%$).
Inception takes images as $299\times299\times3$ dimensional vectors.

\section{Our Approach}

We now turn to our approach for constructing adversarial examples. To begin,
we rely on the initial formulation of adversarial examples
\cite{szegedy2013intriguing} and
formally define the problem of finding an adversarial instance for an image $x$ as follows:
\begin{align*}
  \text{minimize } \;& \mathcal{D}(x, x+\delta)\\
  \text{such that } \;& C(x+\delta) = t \\
    & x+\delta \in [0,1]^n
\end{align*}
where $x$ is fixed, and the goal is to find $\delta$ that minimizes
$\mathcal{D}(x, x+\delta)$.
That is, we want to find some small change $\delta$ that we can make to an
image $x$ that will change its classification, but so that the result is still
a valid image.
Here $\mathcal{D}$ is some distance metric;
for us, it will be either $L_0$, $L_2$, or $L_\infty$ as discussed earlier.

We solve this problem by formulating it as an appropriate optimization
instance that can be solved by existing optimization algorithms.
There are many possible ways to do this; we explore the space of formulations
and empirically identify which ones lead to the most effective attacks.

\subsection{Objective Function}
The above formulation is difficult for existing algorithms to solve directly,
as the constraint $C(x+\delta) = t$ is highly non-linear.
Therefore, we express it in a different form that is better suited for optimization.
We define an objective function $f$ such that
$C(x+\delta) = t$ if and only if $f(x+\delta) \le 0$.
There are many possible choices for $f$:
\begin{align*}
f_{1}(x') &= -\loss_{F,t}(x') + 1\\ 
f_{2}(x') &= (\max_{i \ne t}( F(x')_i) - F(x')_t)^+\\
f_{3}(x') &= \text{softplus}(\max_{i \ne t}(F(x')_i) - F(x')_t)-\log(2)\\
f_{4}(x') &= (0.5 - F(x')_t)^+\\
f_{5}(x') &= -\log(2 F(x')_t - 2)\\ 
f_{6}(x') &= (\max_{i \ne t}( Z(x')_i) - Z(x')_t)^+\\
f_{7}(x') &= \text{softplus}(\max_{i \ne t}(Z(x')_i) - Z(x')_t)-\log(2)
\end{align*}
where $s$ is the correct classification, $(e)^+$ is short-hand for
$\max(e,0)$, $\text{softplus}(x) = \log(1+\exp(x))$, and
$\loss_{F,s}(x)$ is the cross entropy loss for $x$.

Notice that we have adjusted some of the above formula by adding a constant;
we have done this only so that the function respects our definition. This
does not impact the final result, as it just scales the minimization function.

Now, instead of formulating the problem as
\begin{align*}
  \text{minimize } \;& \mathcal{D}(x, x+\delta)\\
  \text{such that } \;& f(x+\delta) \le 0\\
    & x+\delta \in [0,1]^n
\end{align*}
we use the alternative formulation:
\begin{align*}
  \text{minimize } \;& \mathcal{D}(x, x+\delta) + c \cdot f(x+\delta)\\
  \text{such that } \;& x+\delta \in [0,1]^n
\end{align*}
where $c>0$ is a suitably chosen constant.
These two are equivalent, in the sense that there exists $c>0$ such
that the optimal solution to the latter matches the optimal solution
to the former.
After instantiating the distance metric $\mathcal{D}$ with an $l_p$ norm,
the problem becomes: given $x$, find $\delta$ that solves
\begin{align*}
  \text{minimize } \;& \|\delta\|_p + c \cdot f(x+\delta)\\
  \text{such that } \;& x+\delta \in [0,1]^n
\end{align*}

\vspace{3mm}
\emph{Choosing the constant c.}

Empirically, we have found that often the best way to choose $c$ is to
use the smallest value of $c$ for which the resulting solution $x^*$ has
$f(x^*) \le 0$. This causes gradient descent to minimize both of the
terms simultaneously instead of picking only one to optimize over first.

\begin{figure}
  \centering
  \includegraphics[scale=.66]{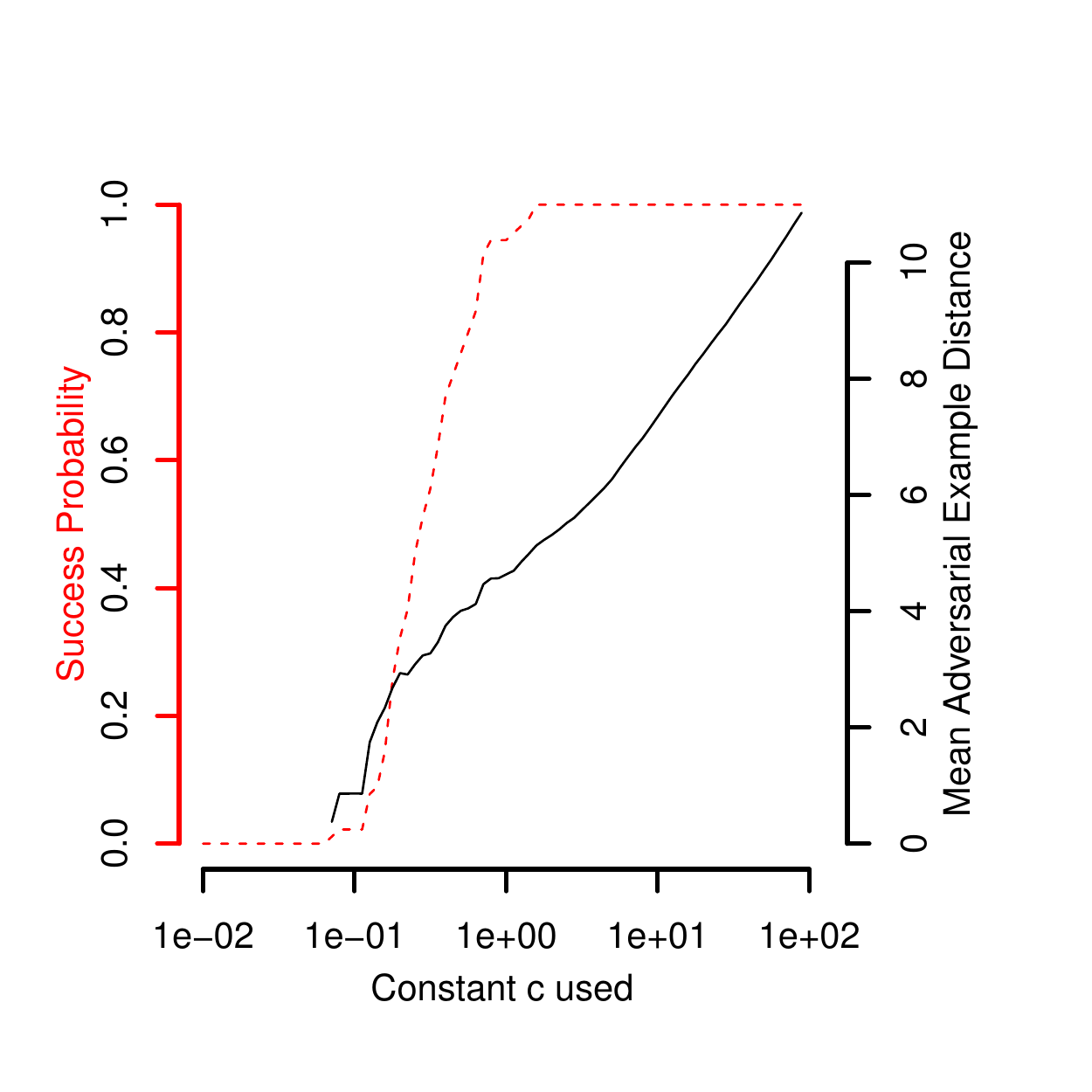}
  \caption{Sensitivity on the constant $c$.
    We plot the $L_2$ distance of the adversarial example computed
    by gradient descent as a function
    of $c$, for objective function $f_6$.
    When $c<.1$, the attack rarely succeeds. After $c>1$, the attack becomes less
    effective, but always succeeds.}
  \label{fig:sensitivity}
\end{figure}

We verify this by running our $f_6$ formulation (which we found
most effective) for values of $c$ spaced
uniformly (on a log scale) from $c=0.01$ to $c=100$ on the MNIST
dataset. We plot this
line in Figure~\ref{fig:sensitivity}. \footnote{The corresponding figures
  for other objective functions are similar; we omit them for brevity.}

Further, we have found that if choose the smallest $c$ such that $f(x^*) \le 0$,
the solution is within $5\%$ of optimal $70\%$ of the time,
and within $30\%$ of optimal $98\%$ of the time, where ``optimal''
refers to the solution found using the best value of $c$.
Therefore, in our implementations we use modified binary search to choose
$c$.

\begin{table*}[t]
\centering
\begin{tabular}{lr@{\quad}r@{\quad}r@{\quad}r@{\quad}r@{\quad}r||@{\quad}r@{\quad}r@{\quad}r@{\quad}r@{\quad}r@{\quad}r||@{\quad}r@{\quad}r@{\quad}r@{\quad}r@{\quad}r@{\quad}r}
  \toprule
 & \multicolumn{6}{c}{\textbf{Best Case}} &
 \multicolumn{6}{c}{\textbf{Average Case}} &
 \multicolumn{6}{c}{\textbf{Worst Case}}  \\
  
 \toprule
 & \multicolumn{2}{c}{Change of} &
 \multicolumn{2}{c}{Clipped} &
 \multicolumn{2}{c}{Projected} 
 & \multicolumn{2}{c}{Change of} &
 \multicolumn{2}{c}{Clipped} &
 \multicolumn{2}{c}{Projected} 
 & \multicolumn{2}{c}{Change of} &
 \multicolumn{2}{c}{Clipped} &
 \multicolumn{2}{c}{Projected} \\
 & \multicolumn{2}{c}{Variable} &
 \multicolumn{2}{c}{Descent} &
 \multicolumn{2}{c}{Descent} & 
  \multicolumn{2}{c}{Variable} &
 \multicolumn{2}{c}{Descent} &
 \multicolumn{2}{c}{Descent} & 
  \multicolumn{2}{c}{Variable} &
 \multicolumn{2}{c}{Descent} &
 \multicolumn{2}{c}{Descent} \\

 & mean & prob & mean & prob & mean & prob &
   mean & prob & mean & prob & mean & prob &
   mean & prob & mean & prob & mean & prob  \\
 \midrule

 $f_1$ & 2.46 & 100$\%$ & 2.93 & 100$\%$ & 2.31 & 100$\%$ & 4.35 & 100$\%$ & 5.21 & 100$\%$ & 4.11 & 100$\%$ & 7.76 & 100$\%$ & 9.48 & 100$\%$ & 7.37 & 100$\%$ \\
 $f_2$ & 4.55 & 80$\%$ & 3.97 & 83$\%$ & 3.49 & 83$\%$ & 3.22 & 44$\%$ & 8.99 & 63$\%$ & 15.06 & 74$\%$ & 2.93 & 18$\%$ & 10.22 & 40$\%$ & 18.90 & 53$\%$ \\
 $f_3$ & 4.54 & 77$\%$ & 4.07 & 81$\%$ & 3.76 & 82$\%$ & 3.47 & 44$\%$ & 9.55 & 63$\%$ & 15.84 & 74$\%$ & 3.09 & 17$\%$ & 11.91 & 41$\%$ & 24.01 & 59$\%$ \\
 $f_4$ & 5.01 & 86$\%$ & 6.52 & 100$\%$ & 7.53 & 100$\%$ & 4.03 & 55$\%$ & 7.49 & 71$\%$ & 7.60 & 71$\%$ & 3.55 & 24$\%$ & 4.25 & 35$\%$ & 4.10 & 35$\%$ \\
 $f_5$ & 1.97 & 100$\%$ & 2.20 & 100$\%$ & 1.94 & 100$\%$ & 3.58 & 100$\%$ & 4.20 & 100$\%$ & 3.47 & 100$\%$ & 6.42 & 100$\%$ & 7.86 & 100$\%$ & 6.12 & 100$\%$ \\
 $f_6$ & 1.94 & 100$\%$ & 2.18 & 100$\%$ & 1.95 & 100$\%$ & 3.47 & 100$\%$ & 4.11 & 100$\%$ & 3.41 & 100$\%$ & 6.03 & 100$\%$ & 7.50 & 100$\%$ & 5.89 & 100$\%$ \\
 $f_7$ & 1.96 & 100$\%$ & 2.21 & 100$\%$ & 1.94 & 100$\%$ & 3.53 & 100$\%$ & 4.14 & 100$\%$ & 3.43 & 100$\%$ & 6.20 & 100$\%$ & 7.57 & 100$\%$ & 5.94 & 100$\%$ \\
 
 \bottomrule
\end{tabular}
\vskip 0.1in
\caption{Evaluation of all combinations of one of the seven possible objective functions with one of the three
  box constraint encodings.
  We show the average $L_2$ distortion,
  the standard deviation, and the
  success probability (fraction of instances for which an adversarial
  example can be found). Evaluated on 1000 random instances. When the success is not $100\%$,
  mean is for successful attacks only.}
  \label{tbl:lossfunctions}
\end{table*}

\subsection{Box constraints}
To ensure the modification yields a valid image,
we have a constraint on $\delta$: we must
have $0 \le x_i+\delta_i \le 1$ for all $i$.
In the optimization literature, this is known as a ``box constraint.''
Previous work uses a particular optimization algorithm, L-BFGS-B,
which supports box constraints natively.

We investigate three different methods of approaching this problem.
\begin{enumerate}
\item \emph{Projected gradient descent} performs one step of standard
  gradient descent, and then clips all the coordinates to be within the box.

This approach can work
poorly for gradient descent approaches that have a complicated update step (for example,
those with momentum): when we clip the actual $x_i$, we
unexpectedly change the input to the next
iteration of the algorithm.

\item \emph{Clipped gradient descent} does not clip $x_i$ on each iteration;
rather, it incorporates the clipping into the objective function to be minimized.
In other words, we replace $f(x+\delta)$ with
$f(\min(\max(x+\delta,0), 1))$, with the min and max taken component-wise.

While solving the main issue with projected gradient descent, clipping
introduces a new problem: the algorithm can get stuck in a flat spot where
it has increased some component $x_i$ to be substantially larger than the maximum
allowed.
When this happens, the partial derivative becomes zero, so even if some
improvement is possible by later reducing $x_i$, gradient descent has no
way to detect this.

\item \emph{Change of variables} introduces a new variable $w$ and
  instead of optimizing over
  the variable $\delta$ defined above, we apply a change-of-variables and
  optimize over $w$, setting
  $$\delta_i= \frac12 (\tanh(w_i) + 1) -x_i.$$
  Since $-1 \le \tanh(w_i) \le 1$, it follows that $0 \le x_i+\delta_i \le 1$,
  so the solution will automatically be valid.
  \footnote{Instead of scaling by $\frac12$ we scale by $\frac12+\epsilon$ to
    avoid dividing by zero.}

We can think of this approach
as a smoothing of clipped gradient descent that eliminates the problem
of getting stuck in extreme regions.




\end{enumerate}

These methods allow us to use
other optimization algorithms that don't natively support box constraints.
We use the Adam \cite{kingma2014adam} optimizer almost exclusively, as we have found it to
be the most effective at quickly finding adversarial examples.
We tried three solvers ---
standard gradient descent, gradient descent with momentum, and Adam
--- and all three produced identical-quality solutions.
However, Adam converges substantially more quickly than the others.

\subsection{Evaluation of approaches}
For each possible objective function $f(\cdot)$ and method to enforce
the box constraint, we evaluate the quality of the adversarial examples
found.

To choose the optimal $c$, we perform $20$ iterations of binary search
over $c$.  For each selected value of $c$, we run $10,000$ iterations
of gradient descent with the Adam optimizer.
\footnote{Adam converges to $95\%$ of optimum within $1,000$ iterations
  $92\%$ of the time. For completeness we run it for $10,000$ iterations at each step.}

The results of this analysis are in Table~\ref{tbl:lossfunctions}.
We evaluate the quality of the adversarial examples found
on the MNIST and CIFAR datasets. The relative ordering of each objective function
is identical between the two datasets, so for brevity
we report only results for MNIST.

There is a factor of three difference in quality between the best objective function
and the worst.
The choice of method for handling box constraints does not impact the quality of
results as significantly for the best minimization functions.

In fact, the worst performing objective function, cross entropy loss,
is the approach that was most suggested in the literature previously
\cite{szegedy2013intriguing,shaham2015understanding}.

\vspace{3mm}
\emph{Why are some loss functions better than others?}
When $c=0$, gradient descent will not make any move away from the initial
image.
However, a large $c$ often causes the initial steps of
gradient descent to perform in an overly-greedy manner, only traveling in the
direction which can most easily reduce $f$ and ignoring the $\mathcal{D}$
loss --- thus causing gradient descent to find sub-optimal solutions.

This means that for
loss function $f_1$ and $f_4$, there is no good constant $c$ that is
useful throughout the duration of the gradient descent search.
Since the constant $c$ weights the relative importance of the distance term
and the loss term, in order for a fixed constant $c$ to be useful, the
relative value of these two terms should remain approximately equal. This is
not the case for these two loss functions.

To explain why this is the case, we will have to take a side discussion to
analyze how adversarial examples exist. Consider a valid input $x$ and an
adversarial example $x'$ on a network.

What does it look like as we linearly interpolate from $x$ to $x'$? That is,
let $y = \alpha x + (1-\alpha)x'$ for $\alpha \in [0,1]$. It turns out the
value of $Z(\cdot)_t$ is mostly linear from the input to the adversarial example,
and therefore the $F(\cdot)_t$ is a logistic. We verify this fact empirically
by constructing adversarial examples on the first $1,000$ test images on both the
MNIST and CIFAR dataset with our approach, and find the Pearson correlation
coefficient $r>.9$.

Given this, consider loss function $f_4$ (the argument for $f_1$ is similar).
In order for the gradient descent attack to make any change initially, the
constant $c$ will have to be large enough that
\begin{equation*}
  \epsilon < c (f_1(x+\epsilon)-f_1(x))
\end{equation*}
or, as $\epsilon \to 0$,
\begin{equation*}
  1/c < \left|\nabla f_1(x)\right|
\end{equation*}
implying that $c$ must be larger than the inverse of the  gradient to make progress, but the
gradient of $f_1$ is identical to $F(\cdot)_t$ so will be tiny around the
initial image, meaning $c$ will have to be extremely large.

However, as soon as we leave the immediate vicinity of the initial image,
the gradient of $\nabla f_1(x+\delta)$ increases at an exponential rate, making
the large constant $c$ cause gradient descent to perform in an overly greedy manner.

We verify all of this theory empirically.
When we run our attack trying constants chosen from $10^{-10}$ to $10^{10}$
the average constant for loss function $f_4$ was $10^6$.

The average gradient of the loss function
$f_1$ around the valid image is $2^{-20}$ but $2^{-1}$ at the closest adversarial example.
This means $c$ is a million times larger than it has to be, causing the loss
function $f_4$ and $f_1$ to perform worse than any of the others.

\subsection{Discretization}
We model pixel intensities as a (continuous) real number
in the range $[0,1]$.
However, in a valid image, each pixel intensity must be a (discrete)
integer in the range $\{0,1,\dots,255\}$.
This additional requirement is not captured in our formulation.
In practice, we ignore the integrality constraints, solve the continuous
optimization problem, and then round to the nearest integer:
the intensity of the $i$th pixel becomes $\lfloor 255(x_i+\delta_i) \rceil$.

This rounding will slightly degrade the quality of the adversarial
example. If we need to restore the attack quality, we perform greedy
search on the lattice defined by the discrete solutions by changing one
pixel value at a time. This greedy search never failed for any of our attacks.

Prior work has largely ignored the integrality constraints.%
\footnote{One exception: The JSMA attack  \cite{papernot2016limitations} handles
this by only setting the output value to either 0 or 255.}
For instance, when using the fast gradient sign attack with
$\epsilon=0.1$ (i.e., changing pixel values by $10\%$), discretization
rarely affects the success rate of the attack.
In contrast, in our work, we are able to find attacks that make
much smaller changes to the images, so discretization effects cannot
be ignored.
We take care to always generate valid images; when reporting the
success rate of our attacks, they always are for attacks that include
the discretization post-processing.

\section{Our Three Attacks} 

\begin{figure}
  \hspace{1cm}
      \begin{tabular}{llllllllll}
        \multicolumn{10}{c}{Target Classification ($L_2$)} \\
        0\, & 1\, & 2\, & 3\, & 4\,\, & 5\, & 6\, & 7\, & 8\, & 9\, \\
  \end{tabular}  \\
  {\rotatebox[origin=l]{90}{
      \begin{tabular}{llllllllll}
        \multicolumn{10}{c}{Source Classification} \\
        9\, & 8\, & 7\, & 6\, & 5\,\, & 4\, & 3\, & 2\, & 1\, & 0\, \\
  \end{tabular}}}
  \centering
  \includegraphics[scale=0.125]{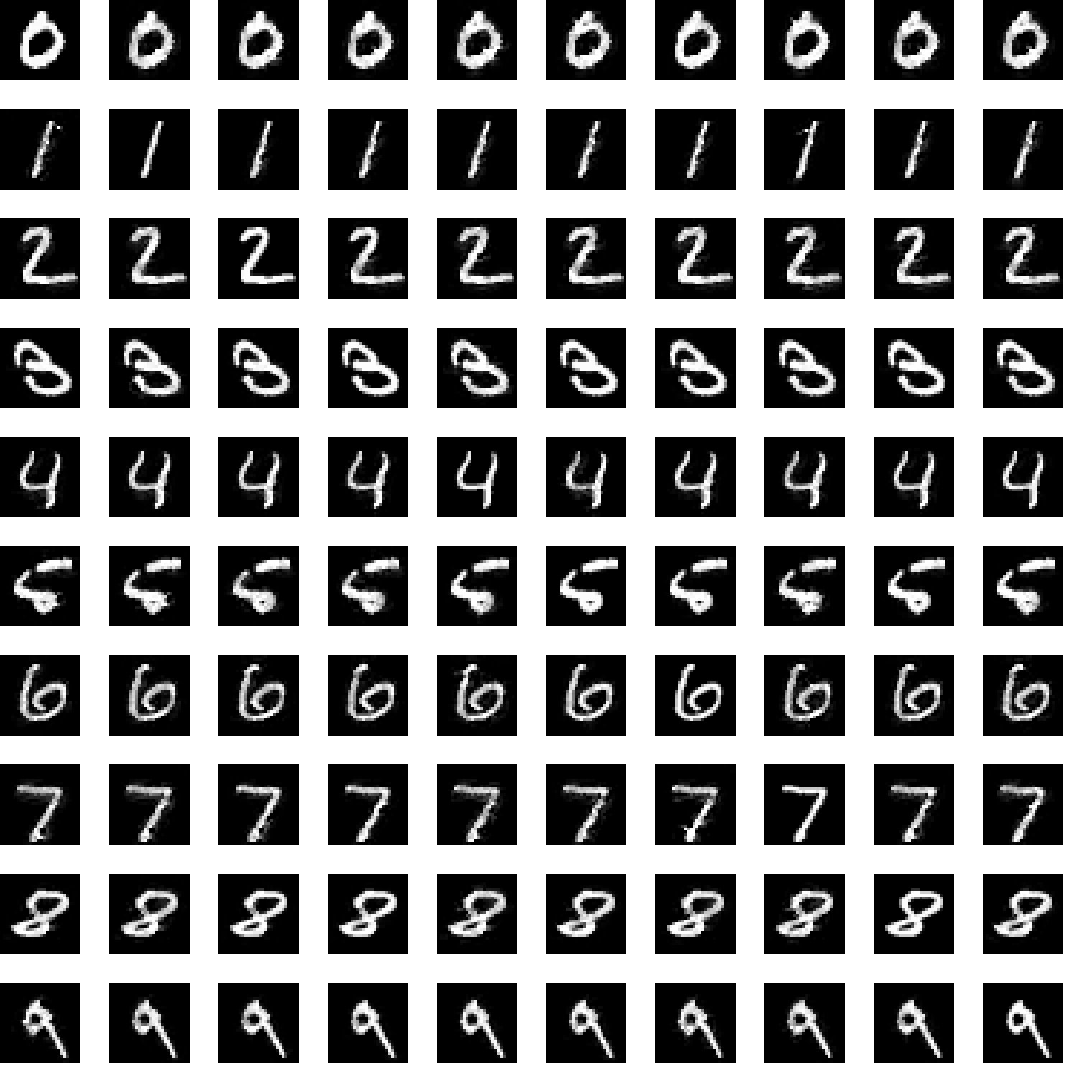}
  \caption{Our $L_2$ adversary applied to the MNIST dataset performing a targeted attack
    for every source/target pair. Each digit is the first image in the dataset with
    that label.}
  \label{fig:mnistl2}
\end{figure}

\begin{figure}
  \hspace{1cm}
      \begin{tabular}{llllllllll}
        \multicolumn{10}{c}{Target Classification ($L_0$)} \\
        0\, & 1\, & 2\, & 3\, & 4\,\, & 5\, & 6\, & 7\, & 8\, & 9\, \\
  \end{tabular}  \\
  {\rotatebox[origin=l]{90}{
      \begin{tabular}{llllllllll}
        \multicolumn{10}{c}{Source Classification} \\
        9\, & 8\, & 7\, & 6\, & 5\,\, & 4\, & 3\, & 2\, & 1\, & 0\, \\
  \end{tabular}}}
  \centering
  \includegraphics[scale=0.125]{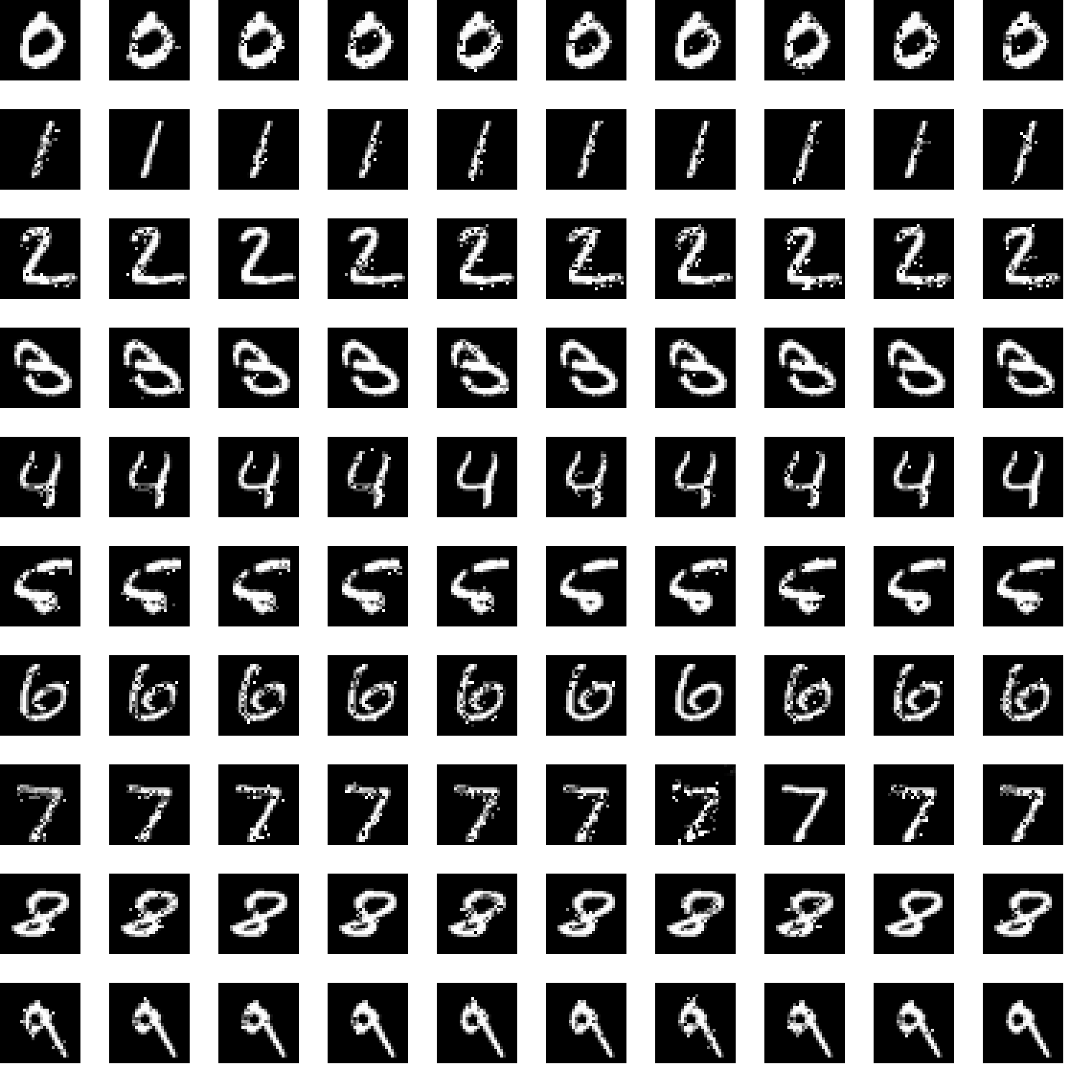}
  \caption{Our $L_0$ adversary applied to the MNIST dataset performing a targeted attack
    for every source/target pair. Each digit is the first image in the dataset with
    that label.}
  \label{fig:mnistl0}
\end{figure}

\subsection{Our $L_2$ Attack} 
Putting these ideas together, we obtain a method for finding
adversarial examples that will have low distortion in the $L_2$ metric.
Given $x$, we choose a target class $t$ (such that we have $t \ne C^*(x)$)
and then search for $w$ that solves
\begin{equation*}
  \text{minimize } \; \|\frac12 (\tanh(w) + 1) -x\|_2^2 + c \cdot f(\frac12 (\tanh(w) + 1)\\
\end{equation*}
with $f$ defined as
\begin{equation*}
f(x') = \max(\max \{ Z(x')_i : i \ne t\} - Z(x')_t, -\kappa).
\end{equation*}
This $f$ is based on the best objective function found earlier,
modified slightly so that we can control the confidence with which the misclassification
occurs by adjusting $\kappa$.
The parameter $\kappa$ encourages the solver to find an adversarial
instance $x'$ that will be classified as class $t$ with high confidence.
We set $\kappa=0$ for our attacks but we note here that a side benefit
of this formulation is it allows one to control for the desired confidence.
This is discussed further in Section~\ref{sec:transfer}.

Figure~\ref{fig:mnistl2} shows this attack applied to our MNIST model for
each source digit and target digit. Almost all attacks are visually
indistinguishable from the original digit.

A comparable figure (Figure~\ref{fig:cifarl2}) for CIFAR is in the appendix. 
No attack is visually distinguishable from the baseline image.

\vspace{3mm}
\emph{Multiple starting-point gradient descent.}
The main problem with gradient descent is that its greedy search is not
guaranteed to find the optimal solution and can become stuck in a local
minimum. To remedy this, we pick multiple
random starting points close to the original image and run
gradient descent from each of those
points for a fixed number of iterations.
We randomly sample points uniformly from the ball of radius $r$, where $r$
is the closest adversarial example found so far.
Starting from multiple starting points reduces the likelihood that gradient descent gets
stuck in a bad local minimum.


\subsection{Our $L_0$ Attack} 
The $L_0$ distance metric is non-differentiable and therefore is ill-suited for
standard gradient descent.
Instead, we use an iterative algorithm that, in each iteration, identifies some
pixels that don't have much effect on the classifier output and then fixes
those pixels, so their value will never be changed.
The set of fixed pixels grows in each iteration until we have, by
process of elimination, identified a minimal (but possibly not minimum)
subset of pixels that can be
modified to generate an adversarial example.
In each iteration, we use our $L_2$ attack
to identify which pixels are unimportant.



In more detail, on each iteration, we call the $L_2$ adversary, restricted
to only modify the pixels in the allowed set.
Let $\delta$ be the solution returned from the $L_2$ adversary on input image $x$,
so that $x+\delta$ is an adversarial example.
We compute $g = \nabla f(x+\delta)$ (the gradient of the objective function,
evaluated at the adversarial instance).
We then select the pixel $i = \arg\min_i g_i \cdot \delta_i$
and fix $i$, i.e., remove $i$ from the allowed set.%
\footnote{Selecting the index $i$ that minimizes $\delta_i$ is simpler,
but it yields results with $1.5\times$ higher $L_0$ distortion.}
The intuition is that $g_i \cdot \delta_i$ tells us how much
reduction to $f(\cdot)$ we obtain from the $i$th pixel of the image,
when moving from $x$ to $x+\delta$: $g_i$ tells us how much reduction in $f$
we obtain, per unit change to the $i$th pixel, and we multiply this by
how much the $i$th pixel has changed.
This process repeats until the $L_2$ adversary fails to find an adversarial example.

There is one final detail required to achieve strong results: choosing a constant
$c$ to use for the $L_2$ adversary. To do this, we initially
set $c$ to a very low value (e.g., $10^{-4}$). We then run our $L_2$ adversary
at this $c$-value. If it fails, we double $c$ and try again, until it is successful.
We abort the search if $c$ exceeds a fixed threshold (e.g., $10^{10}$).

JSMA \emph{grows} a set --- initially empty
--- of pixels that are allowed to be changed and sets the pixels to maximize the
total loss. In contrast, our attack \emph{shrinks} the set of pixels
 --- initially containing every pixel --- that are allowed to be changed.

Our algorithm is significantly more effective than JSMA (see
Section~\ref{sec:eval} for an evaluation).
It is also efficient: we introduce optimizations that make it about as
fast as our $L_2$ attack with a single starting point on MNIST and CIFAR; it is
substantially slower on ImageNet.
Instead of starting gradient descent in each iteration from the initial image,
we start the gradient descent from the solution found on the previous iteration
(``warm-start'').
This dramatically reduces the number of rounds of gradient descent needed during
each iteration, as the solution with $k$ pixels held constant is often very similar
to the solution with $k+1$ pixels held constant.

Figure~\ref{fig:mnistl0} shows the $L_0$ attack applied to
one digit of each source class, targeting each target class, on the MNIST dataset.
The attacks are visually noticeable, implying the $L_0$ attack is more difficult than
$L_2$. Perhaps the worst case is that of a 7 being made to classify as a 6; interestingly,
this attack for $L_2$ is one of the only visually distinguishable attacks.

A comparable figure (Figure~\ref{fig:cifarl0}) for CIFAR is in the appendix.

\begin{figure}
  \hspace{1cm}
      \begin{tabular}{llllllllll}
        \multicolumn{10}{c}{Target Classification ($L_\infty$)} \\
        0\, & 1\, & 2\, & 3\, & 4\,\, & 5\, & 6\, & 7\, & 8\, & 9\, \\
  \end{tabular}  \\
  {\rotatebox[origin=l]{90}{
      \begin{tabular}{llllllllll}
        \multicolumn{10}{c}{Source Classification} \\
        9\, & 8\, & 7\, & 6\, & 5\,\, & 4\, & 3\, & 2\, & 1\, & 0\, \\
  \end{tabular}}}
  \centering
  \includegraphics[scale=0.125]{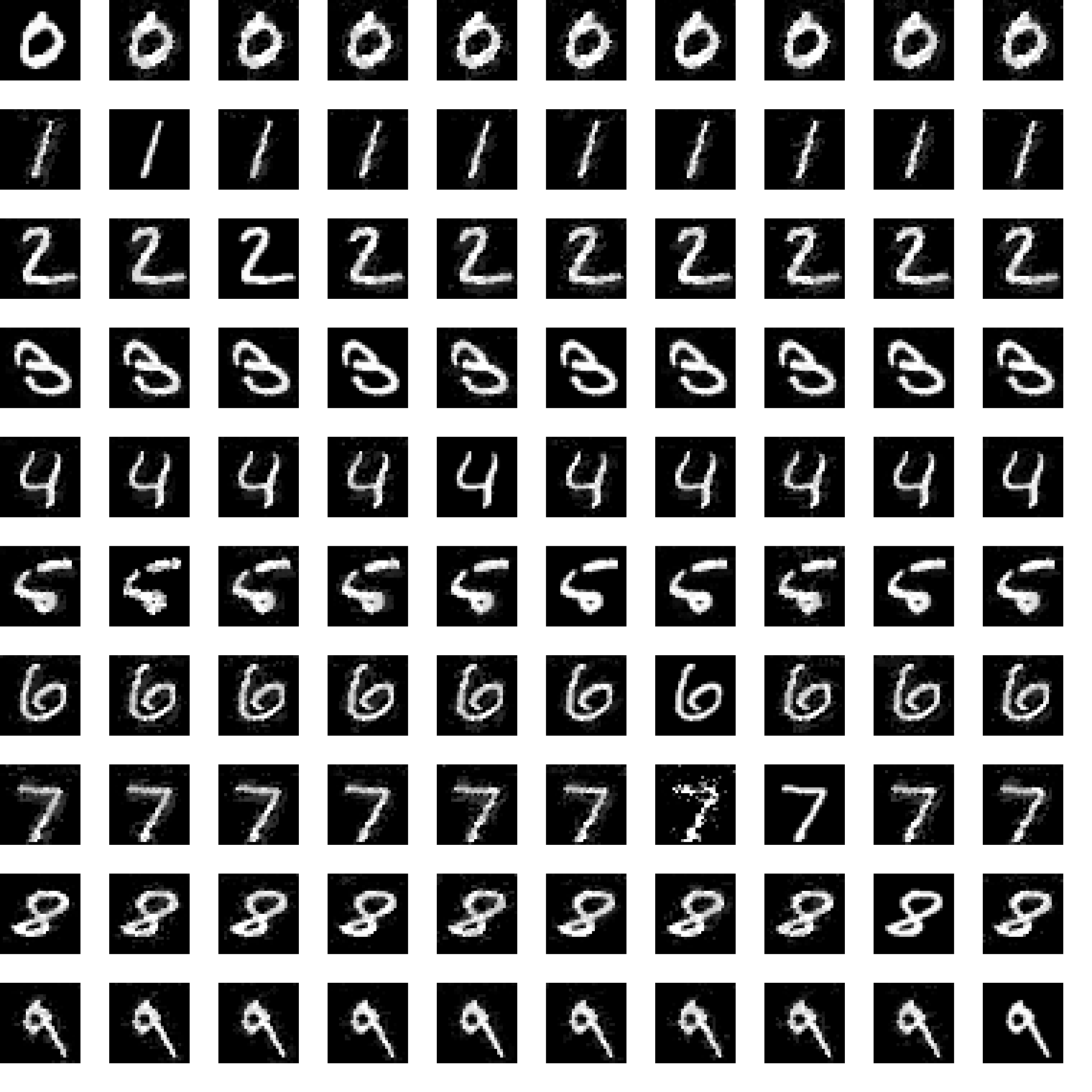}
  \caption{Our $L_\infty$ adversary applied to the MNIST dataset performing a targeted attack
    for every source/target pair. Each digit is the first image in the dataset with
    that label.}
  \label{fig:mnistli}
\end{figure}

\subsection{Our $L_{\infty}$ Attack} 
The $L_{\infty}$ distance metric is not fully differentiable and
standard gradient descent does not perform well for it.
We experimented with naively optimizing
\begin{equation*}
\text{minimize } \; \; c \cdot f(x+\delta) + \|\delta\|_\infty
\end{equation*}
However, we found that gradient descent produces very poor results: the $\|\delta\|_\infty$ term only
penalizes the largest (in absolute value) entry in $\delta$ and has no impact on
any of the other. As such, gradient descent very quickly becomes stuck
oscillating between two suboptimal solutions. Consider a
case where $\delta_i = 0.5$ and $\delta_j = 0.5-\epsilon$.
The $L_{\infty}$ norm will only penalize $\delta_i$, not $\delta_j$,
and $\frac{\partial}{\partial \delta_j} \|\delta\|_\infty$ will be zero
at this point.
Thus, the gradient imposes no penalty for increasing $\delta_j$,
even though it is already large.
On the next iteration we might move to a position where $\delta_j$ is
slightly larger than $\delta_i$, say $\delta_i = 0.5 - \epsilon'$
and $\delta_j = 0.5 + \epsilon''$, a mirror image of where we started.
In other words, gradient descent may oscillate back and forth across
the line $\delta_i=\delta_j=0.5$,
making it nearly impossible to make progress.

We resolve this issue using an iterative attack.
We replace the $L_2$ term in
the objective function with a penalty for any terms that exceed $\tau$
(initially $1$, decreasing in each iteration).
This prevents oscillation, as this loss term penalizes all large values
simultaneously. Specifically, in each iteration we solve
\begin{equation*}
\text{minimize } \;\;
c \cdot f(x+\delta) + \cdot \sum_i \left[(\delta_i-\tau)^+\right]
\end{equation*}
After each iteration, if $\delta_i < \tau$ for all $i$, we
reduce $\tau$ by a factor of 0.9 and repeat; otherwise, we terminate
the search.

Again we must choose a good constant
$c$ to use for the $L_\infty$ adversary. We take the same approach as we do for the $L_0$
attack: initially
set $c$ to a very low value and run the $L_\infty$ adversary
at this $c$-value. If it fails, we double $c$ and try again, until it is successful.
We abort the search if $c$ exceeds a fixed threshold.

Using ``warm-start'' for gradient descent in each iteration,
this algorithm is about as fast as our $L_2$ algorithm (with a single
starting point).

Figure~\ref{fig:mnistli} shows the $L_\infty$ attack applied to
one digit of each source class, targeting each target class, on the MNSIT dataset.
While most differences are not visually noticeable, a few are. Again, the worst case is that
of a 7 being made to classify as a 6.

A comparable figure (Figure~\ref{fig:cifarli}) for CIFAR is in the appendix. No attack
is visually distinguishable from the baseline image.

\section{Attack Evaluation}

\begin{table*}[t]
\centering
\begin{tabular}{ld{3.2}rd{3.2}r||d{3.2}rd{3.2}r||d{3.2}rd{3.2}r}
  \toprule
 & \multicolumn{4}{c}{\textbf{Best Case}} &
 \multicolumn{4}{c}{\textbf{Average Case}} &
 \multicolumn{4}{c}{\textbf{Worst Case}}  \\
  
 \toprule
 & \multicolumn{2}{c}{MNIST} &
 \multicolumn{2}{c}{CIFAR}
 & \multicolumn{2}{c}{MNIST} &
 \multicolumn{2}{c}{CIFAR}
 & \multicolumn{2}{c}{MNIST} &
 \multicolumn{2}{c}{CIFAR} \\

 & \multicolumn{1}{c}{mean} & prob & \multicolumn{1}{c}{mean} & prob &
   \multicolumn{1}{c}{mean} & prob & \multicolumn{1}{c}{mean} & prob &
   \multicolumn{1}{c}{mean} & prob & \multicolumn{1}{c}{mean} & prob \\
 \midrule

 Our $L_0$ & 8.5 & $100\%$ & 5.9 & $100\%$
 & 16 & $100\%$ & 13 & $100\%$
  & 33 & $100\%$ & 24 & $100\%$ \\

  JSMA-Z & 20 & $100\%$ & 20 & $100\%$
 & 56 & $100\%$ & 58 & $100\%$
  & 180 & $98\%$ & 150 & $100\%$ \\

  JSMA-F & 17 & $100\%$ & 25 & $100\%$
 & 45 & $100\%$ & 110 & $100\%$
  & 100 & $100\%$ & 240 & $100\%$ \\
  
 \midrule
 
 Our $L_2$  & 1.36 & $100\%$ & 0.17 & $100\%$
 & 1.76 & $100\%$ & 0.33 & $100\%$ 
  & 2.60 & $100\%$ & 0.51 & $100\%$ \\

  Deepfool & 2.11 & $100\%$ & 0.85 & $100\%$
 & - & - & - & -
  & - & - & - & - \\

 \midrule
 
 Our $L_\infty$  & 0.13 & $100\%$ & 0.0092 & $100\%$ 
 & 0.16 & $100\%$ & 0.013 & $100\%$ 
  & 0.23 & $100\%$ & 0.019 & $100\%$ \\

  Fast Gradient Sign & 0.22 & $100\%$ & 0.015 & $99\%$
 & 0.26 & $42\%$ & 0.029 & $51\%$
  & - & $0\%$ & 0.34 & $1\%$ \\

  Iterative Gradient Sign & 0.14 & $100\%$ & 0.0078 & $100\%$
 & 0.19 & $100\%$ & 0.014 & $100\%$
  & 0.26 & $100\%$ & 0.023 & $100\%$ \\
  
 \bottomrule
\end{tabular}
\vskip 0.1in
\caption{Comparison of the three variants of targeted attack to previous work for our
  MNIST and CIFAR models. When success rate is not $100\%$, the mean is only over successes. }
  \label{tbl:eval}
\end{table*}

\begin{table}[t]
\centering
\begin{tabular}{lrr||rr||rr}
  \toprule
 & \multicolumn{2}{c}{\textbf{Untargeted}} &
 \multicolumn{2}{c}{\textbf{Average Case}} &
 \multicolumn{2}{c}{\textbf{Least Likely}}  \\
  
 \toprule

 & mean & prob &
   mean & prob &
   mean & prob \\
 \midrule

 Our $L_0$ & 48 & $100\%$ 
 & 410 & $100\%$ 
  & 5200 & $100\%$ \\

  JSMA-Z & - & $0\%$
 & - & $0\%$ 
  & - & $0\%$ \\

  JSMA-F & - & $0\%$
 & - & $0\%$ 
  & - & $0\%$ \\
  
 \midrule
 
 Our $L_2$  & 0.32 & $100\%$ 
 & 0.96 & $100\%$ 
  & 2.22 & $100\%$ \\

  Deepfool & 0.91 & $100\%$ 
 & - & - 
  & - & - \\

 \midrule
 
 Our $L_\infty$  & 0.004 & $100\%$
 & 0.006 & $100\%$ 
  & 0.01 & $100\%$ \\

  FGS & 0.004 & $100\%$ 
 & 0.064 & $2\%$ 
  & - & $0\%$ \\

  IGS & 0.004 & $100\%$ 
 & 0.01 & $99\%$ 
  & 0.03 & $98\%$ \\
  
 \bottomrule
\end{tabular}
\vskip 0.1in
\caption{Comparison of the three variants of targeted attack to previous work for the Inception v3
  model on ImageNet. When success rate is not $100\%$, the mean is only over successes.}
  \label{tbl:evalimagenet}
\end{table}

\label{sec:eval}
We compare our targeted attacks to the best results
previously reported in prior publications,
for each of the three distance metrics.

We re-implement Deepfool, fast gradient sign, and iterative gradient sign.
For fast gradient sign, we search over $\epsilon$ to find the smallest distance
that generates an adversarial example; failures is returned if no $\epsilon$
produces the target class. Our iterative gradient sign method is similar:
we search over $\epsilon$ (fixing $\alpha={1 \over 256}$) and return the
smallest successful.

For JSMA we use the implementation
in CleverHans \cite{papernot2016cleverhans} with only slight modification (we
improve performance by $50\times$ with no impact on accuracy).

JSMA is unable to run on ImageNet due to an inherent significant computational
cost: recall that JSMA performs search for a pair of pixels $p,q$ that can be
changed together that make the target class more likely and other classes less
likely. ImageNet represents images as $299\times299\times3$ vectors, so searching
over all pairs of pixels would require $2^{36}$ work on each step of the calculation.
If we remove the search over pairs of pixels, the success of JSMA falls off
dramatically. We therefore report it as failing always on ImageNet.

We report success if the attack produced an adversarial example with the
correct target label, no matter how much change was required. Failure
indicates the case where the attack was entirely unable to succeed.

We evaluate on the first $1,000$ images in the test set on CIFAR and MNSIT.
On ImageNet, we report on  $1,000$ images that were initially classified
correctly by Inception v3 \footnote{Otherwise the best-case attack results would
  appear to succeed extremely often artificially  low due to the relatively low
  top-1 accuracy}. On ImageNet we approximate the best-case and worst-case
results by choosing $100$ target classes ($10\%$) at random.

The results are found in Table~\ref{tbl:eval} for MNIST and CIFAR, and Table~\ref{tbl:evalimagenet}
for ImageNet. \footnote{The complete code to reproduce these
  tables and figures is available online at \url{http://nicholas.carlini.com/code/nn_robust_attacks}.} 

For each distance metric, across all three datasets, our attacks find closer
adversarial examples than the previous state-of-the-art attacks, and
our attacks never fail to find an adversarial example.
Our $L_0$ and $L_2$ attacks find adversarial examples with
$2\times$ to $10\times$ lower distortion than the best previously published
attacks, and succeed with $100\%$ probability.
Our $L_\infty$ attacks are comparable in quality to prior work, but
their success rate is higher.
Our $L_\infty$ attacks on ImageNet are
so successful that we can change the classification of an image to any
desired label by only flipping
the lowest bit of each pixel, a change that would be impossible to detect visually.

As the learning task becomes increasingly
more difficult, the previous attacks produce worse results,
due to the complexity of the model.
In contrast, our attacks perform even
better as the task complexity increases. We have found JSMA is unable
to find targeted $L_0$ adversarial examples on ImageNet, whereas ours is able to with
$100\%$ success.

It is important to realize that the results between models are not directly comparable. For example, even though a $L_0$ adversary must change $10$ times
as many pixels to switch an ImageNet
classification compared to a MNIST classification,
ImageNet has $114\times$ as many pixels and so the \emph{fraction of pixels}
that must change is significantly smaller.


\begin{figure}
  \hspace{1cm}
      \begin{tabular}{llllllllll}
        \multicolumn{10}{c}{Target Classification} \\
        0\, & 1\, & 2\, & 3\, & 4\,\, & 5\, & 6\, & 7\, & 8\, & 9\, \\
  \end{tabular} \\
  {\rotatebox[origin=l]{90}{
      \begin{tabular}{lll}
        \multicolumn{3}{c}{Distance Metric} \\
        $L_\infty$\, & $L_2$\, & $L_0$\\
  \end{tabular}}}
  \centering
  \includegraphics[scale=0.125]{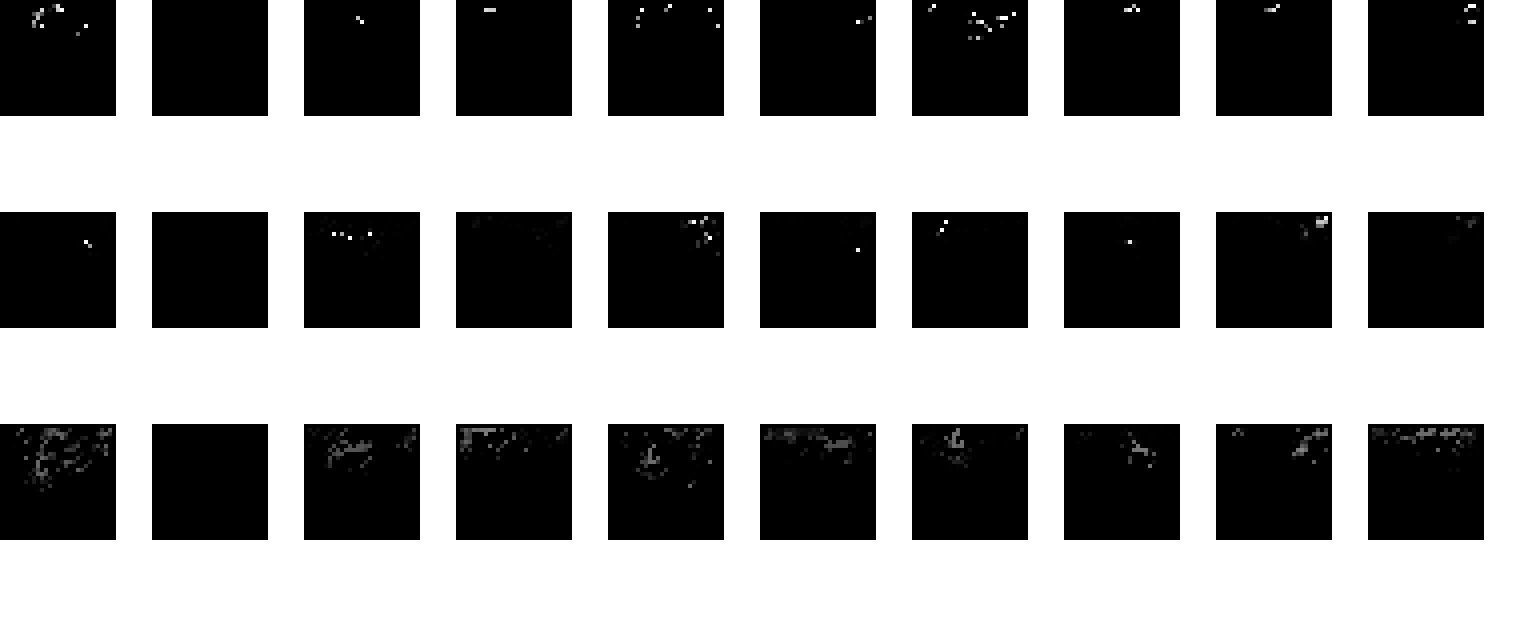}
  \caption{Targeted attacks for each of the 10 MNIST digits where the starting
  image is totally black for each of the three distance metrics.}
  \label{fig:create0}
\end{figure}

\begin{figure}
  \hspace{1cm}
      \begin{tabular}{llllllllll}
        \multicolumn{10}{c}{Target Classification} \\
        0\, & 1\, & 2\, & 3\, & 4\,\, & 5\, & 6\, & 7\, & 8\, & 9\, \\
  \end{tabular} \\
  {\rotatebox[origin=l]{90}{
      \begin{tabular}{lll}
        \multicolumn{3}{c}{Distance Metric} \\
        $L_\infty$\, & $L_2$\, & $L_0$\\
  \end{tabular}}}
  \centering
  \includegraphics[scale=0.125]{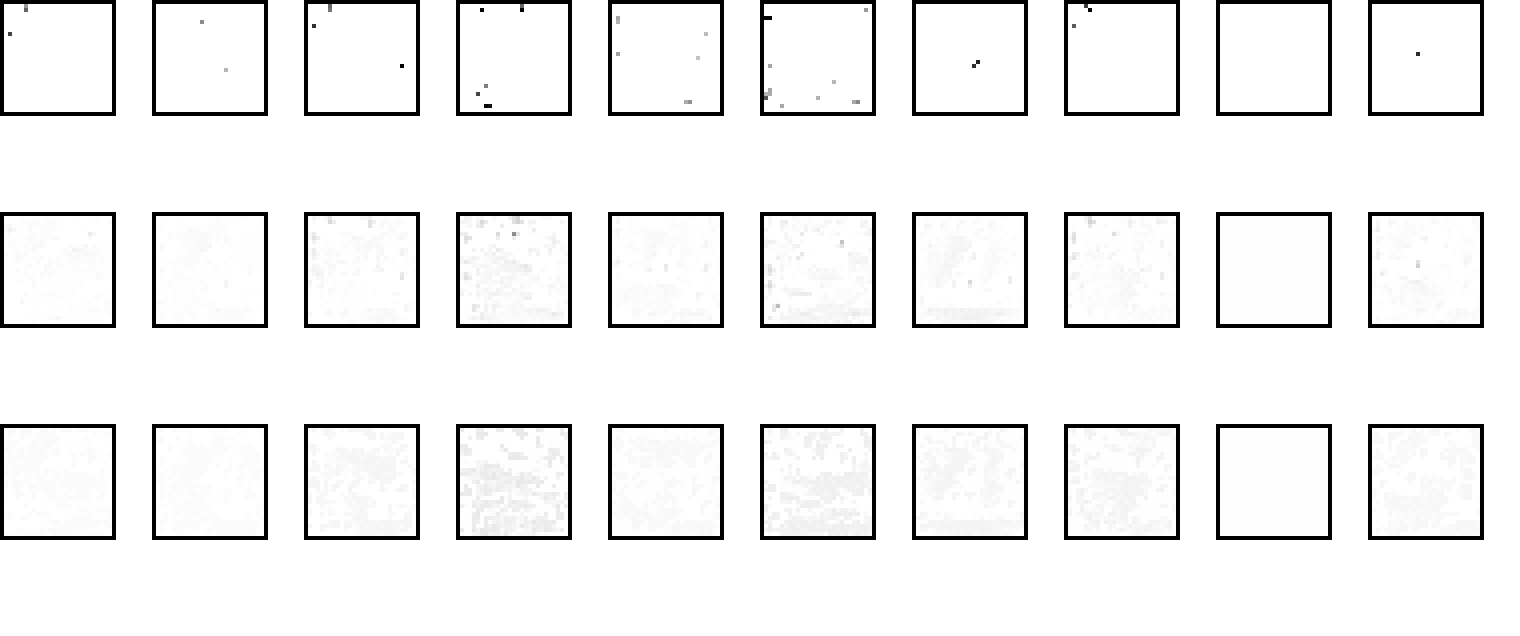}
  \caption{Targeted attacks for each of the 10 MNIST digits where the starting
    image is totally white for each of the three distance metrics.}
  \label{fig:create1}
\end{figure}

\vspace{3mm}
\emph{Generating synthetic digits.}
With our targeted adversary, we can start from \emph{any} image we want and find adversarial
examples of each given target. Using this, in Figure~\ref{fig:create0} we show the
minimum perturbation
to an entirely-black image required to make it classify as each digit, for each of the
distance metrics.

This experiment was performed for the $L_0$ task previously \cite{papernot2016limitations},
however when mounting their attack, ``for classes 0, 2, 3 and 5 one can clearly recognize
the target digit.'' With our more powerful attacks, none of the digits are recognizable.
Figure~\ref{fig:create1} performs the same analysis starting from an all-white image.

Notice that the all-black image requires no change to become a digit $1$ because it
is initially classified as a $1$, and the all-white image requires no change to become
a $8$ because the initial image is already an $8$.

\vspace{3mm}
\emph{Runtime Analysis.}
We believe there are two reasons why one may consider the runtime performance of
adversarial example generation algorithms important: first, to understand if the
performance would be prohibitive for an adversary to actually mount the attacks, and
second, to be used as an inner loop in adversarial re-training \cite{goodfellow2014explaining}.

Comparing the exact runtime of attacks can be misleading. For example, we have
parallelized the implementation of our $L_2$ adversary allowing it to run hundreds
of attacks simultaneously on a GPU, increasing performance from $10\times$ to $100\times$.
However, we did not parallelize our $L_0$ or $L_\infty$ attacks. Similarly,
our implementation of fast gradient sign is parallelized, but JSMA is not. We therefore
refrain from giving exact performance numbers because we believe an unfair comparison
is worse than no comparison.

All of our attacks, and all previous attacks, are plenty efficient to be used by
an adversary. No attack takes longer than a few minutes to run on any given instance.

When compared to $L_0$, our attacks are $2\times-10\times$ slower than our optimized
JSMA algorithm (and significantly faster than the un-optimized version).
Our attacks are typically $10\times-100\times$ slower than previous attacks for
$L_2$ and $L_\infty$, with exception of iterative gradient sign which we are
$10\times$ slower.

\section{Evaluating Defensive Distillation}
\label{sec:distillation}

\emph{Distillation} was initially proposed as an approach to reduce
a large model (the \emph{teacher}) down to a smaller \emph{distilled} model \cite{hinton2015distilling}.
At a high level, distillation works by first training the teacher model on the
training set in a standard manner.
Then, we use the teacher to label each instance in the training set
with soft labels (the output vector from the teacher network).
For example, while the hard label for
an image of a hand-written digit $7$ will say it is classified as a seven, the 
soft labels  might say it has a $80\%$ chance of being a seven and a $20\%$ chance
of being a one.
Then, we train the distilled model on the soft labels from the teacher,
rather than on the hard labels from the training set.
Distillation can potentially increase accuracy on the test set as well as
the rate at which the smaller model learns to predict the hard labels
\cite{hinton2015distilling,melicher2016fast}.

\emph{Defensive distillation} uses distillation in order to increase
the robustness of a neural network, but with two significant
changes. First, both the teacher model and the distilled model are identical
in size --- defensive distillation does not result in smaller models.
Second, and more importantly, defensive distillation uses a large
\emph{distillation temperature} (described below) to force the distilled model
to become more confident in its predictions.

Recall that, the softmax function is the last layer of a neural network.
Defensive distillation modifies the softmax function to also include a
temperature constant $T$:
\begin{equation*}
\softmax(x, T)_i = \frac{e^{x_i/T}}{\sum_j e^{x_j/T}}
\end{equation*}
It is easy to see that $\softmax(x, T) = \softmax(x/T, 1)$. Intuitively, increasing the 
temperature causes a ``softer'' maximum, and decreasing it causes a ``harder'' maximum.
As the limit of the temperature goes to $0$, $\softmax$ approaches $\max$;
as the limit goes to infinity, $\softmax(x)$ approaches a uniform distribution.

Defensive distillation proceeds in four steps:
\begin{enumerate}
\item Train a network, the teacher network, by setting the temperature of the 
softmax to $T$ during the training phase.
\item Compute soft labels by apply the teacher network to each instance in the training
set, again evaluating the softmax at temperature $T$.
\item Train the distilled network (a network with the same shape as the teacher network) 
on the soft labels, using softmax at temperature $T$.
\item Finally, when running the distilled network at test time (to classify new
inputs), use temperature $1$.
\end{enumerate}

\begin{table*}[t]
\centering
\begin{tabular}{ld{3.2}rd{3.2}r||d{3.2}rd{3.2}r||d{3.2}rd{3.2}r}
  \toprule
 & \multicolumn{4}{c}{\textbf{Best Case}} &
 \multicolumn{4}{c}{\textbf{Average Case}} &
 \multicolumn{4}{c}{\textbf{Worst Case}}  \\
  
 \toprule
 & \multicolumn{2}{c}{MNIST} &
 \multicolumn{2}{c}{CIFAR}
 & \multicolumn{2}{c}{MNIST} &
 \multicolumn{2}{c}{CIFAR}
 & \multicolumn{2}{c}{MNIST} &
 \multicolumn{2}{c}{CIFAR} \\

 & \multicolumn{1}{c}{mean} & prob & \multicolumn{1}{c}{mean} & prob &
   \multicolumn{1}{c}{mean} & prob & \multicolumn{1}{c}{mean} & prob &
   \multicolumn{1}{c}{mean} & prob & \multicolumn{1}{c}{mean} & prob \\
 \midrule

 Our $L_0$ & 10 & $100\%$ & 7.4 & $100\%$
 & 19 & $100\%$ & 15 & $100\%$
  & 36 & $100\%$ & 29 & $100\%$ \\

 \midrule
 
 Our $L_2$  & 1.7 & $100\%$ & 0.36 & $100\%$
 & 2.2 & $100\%$ & 0.60 & $100\%$ 
  & 2.9 & $100\%$ & 0.92 & $100\%$ \\

 \midrule
 
 Our $L_\infty$  & 0.14 & $100\%$ & 0.002 & $100\%$ 
 & 0.18 & $100\%$ & 0.023 & $100\%$ 
  & 0.25 & $100\%$ & 0.038 & $100\%$ \\

 \bottomrule
\end{tabular}
\vskip 0.1in
\caption{Comparison of our attacks when applied to defensively
  distilled networks. Compare to Table~\ref{tbl:eval} for undistilled networks.}
  \label{tbl:distillation}
\end{table*}

\subsection{Fragility of existing attacks}
\label{sec:fragility}
We briefly investigate the reason that existing attacks fail on distilled networks,
and find that existing attacks are very fragile and can easily fail to find adversarial
examples even when they exist.

\vspace{3mm}
\emph{L-BFGS and Deepfool} fail due to the fact that the gradient of $F(\cdot)$ is zero
almost always, which prohibits the use of the standard objective function.

When we train a distilled network at temperature $T$ and then test it at temperature $1$,
we effectively cause the inputs to the softmax to become larger by a factor of $T$.
By minimizing the cross entropy during training, the output of the softmax is forced to
be close to $1.0$ for the correct class and $0.0$ for all others.
Since $Z(\cdot)$ is
divided by $T$, the distilled network will learn to make the $Z(\cdot)$ values $T$ times larger than
they otherwise would be.
(Positive values are forced to become about $T$ times larger;
negative values are multiplied by a factor of about $T$ and thus become even more negative.)
Experimentally, we verified this fact: the mean value of the $L_1$ norm of
$Z(\cdot)$ (the logits) on the undistilled
network is $5.8$ with standard deviation $6.4$; on the distilled network (with $T=100$),
the mean is $482$ with standard deviation $457$.

Because the values of $Z(\cdot)$ are 100 times larger, when we test at temperature $1$,
the output of $F$ becomes $\epsilon$ in all components except for the output class which
has confidence $1-9\epsilon$ for some very small $\epsilon$ (for tasks with 10 classes). In fact, in most cases,
$\epsilon$ is so small that the 32-bit floating-point value is rounded to $0$.
For similar reasons, the gradient is so small that it becomes
$0$ when expressed as a 32-bit floating-point value.

This causes the L-BFGS minimization procedure to fail to
make progress and terminate. If instead we run L-BFGS with our stable objective
function identified earlier, rather than the objective function
$\loss_{F,l}(\cdot)$ suggested by
Szegedy \emph{et al.} \cite{szegedy2013intriguing},
L-BFGS does not fail. An alternate approach to fixing the attack would be to
set
\begin{equation*}
  F'(x) = \softmax(Z(x)/T)
\end{equation*}
where $T$ is the distillation temperature chosen. Then minimizing
$\loss_{F',l}(\cdot)$ will not fail, as now the gradients do not vanish due to
floating-point arithmetic rounding.
This clearly demonstrates the fragility of using the loss function as the objective to
minimize.

\vspace{3mm}
\emph{JSMA-F} (whereby we mean the attack uses the output of the
final layer $F(\cdot)$) fails for the same reason that L-BFGS fails:
the output of the $Z(\cdot)$ layer is very large and so softmax becomes essentially a
hard maximum. This is the version of the attack that Papernot \emph{et al.} use to
attack defensive distillation in their paper \cite{distillation}. 

\vspace{3mm}
\emph{JSMA-Z} (the attack that uses the logits)
fails for a completely different reason.
Recall that in the $Z(\cdot)$ version of the attack, we use the input to the softmax
for computing the gradient instead of the final output of the network. This removes
any potential issues with the gradient vanishing, however this introduces new issues.
This version of the attack is introduced by Papernot \emph{et al.} \cite{papernot2016limitations}
but it is not used to attack distillation; we provide here an analysis of why it fails.

Since this attack uses the $Z$ values, it is important to realize the
differences in relative impact. If the smallest input to the softmax
layer is $-100$, then, after the softmax layer, the corresponding output becomes
practically zero. If this input changes from $-100$ to $-90$, the output will
still be practically zero. However, if the largest input to the softmax layer is
$10$, and it changes to $0$, this will have a massive impact on the softmax
output.

Relating this to parameters used in their attack, 
$\alpha$ and $\beta$ represent the size of the change at the input
to the softmax layer.
It is perhaps surprising that JSMA-Z
works on un-distilled networks, as it treats all changes as being of
equal importance, regardless of how much they change the softmax output.
If changing a single pixel would increase the target class by $10$, but also
increase the least likely class by $15$, the attack will not increase that pixel.

Recall that distillation at temperature $T$ causes the value of the logits to
be $T$ times larger.
In effect, this magnifies the sub-optimality noted above
as logits that are extremely unlikely but have slight variation can cause the
attack to refuse to make any changes.

\vspace{3mm}
\emph{Fast Gradient Sign} fails at first for the same reason L-BFGS fails: the
gradients are almost always zero. However, something interesting happens if we
attempt the same division trick and divide the logits by $T$ before feeding them
to the softmax function: distillation still remains effective \cite{papernot2016effectiveness}.
We are unable to explain this phenomenon.

\subsection{Applying Our Attacks}
When we apply our attacks to defensively distilled networks, we find distillation provides
only marginal value. We re-implement defensive distillation on MNIST and CIFAR-10
as described \cite{distillation}
using the same model we used for our evaluation above. We train our distilled model
with temperature $T=100$, the value found to be most effective \cite{distillation}.

Table~\ref{tbl:distillation} shows our attacks when applied to distillation.
All of the
previous attacks fail to find adversarial examples.
In contrast, our attack succeeds with
$100\%$ success probability for each of the three distance metrics.

When compared to Table~\ref{tbl:eval}, distillation has added almost no value:
our $L_0$ and $L_2$ attacks perform slightly worse,
and our $L_\infty$ attack performs approximately equally. All of our attacks
succeed with $100\%$ success.


\subsection{Effect of Temperature}

In the original work, increasing the temperature was found to consistently reduce
attack success rate. On MNIST, this goes from a $91\%$ success rate at $T=1$ to a
$24\%$ success rate for $T=5$ and finally $0.5\%$ success at $T=100$.

We re-implement this experiment with our improved attacks to understand how
the choice of temperature impacts robustness. We train models with the temperature varied
from $t=1$ to $t=100$.

When we re-run our implementation of JSMA, we observe the same
effect: attack success rapidly decreases. However, with our improved $L_2$ attack, we
see no effect of temperature on the mean distance to adversarial examples: the
correlation coefficient is $\rho = -0.05$. This clearly demonstrates the fact that
increasing the distillation temperature does not increase the robustness of the
neural network, it only causes existing attacks to fail more often.

\begin{figure}[t]
  \centering
  \includegraphics[scale=.66667]{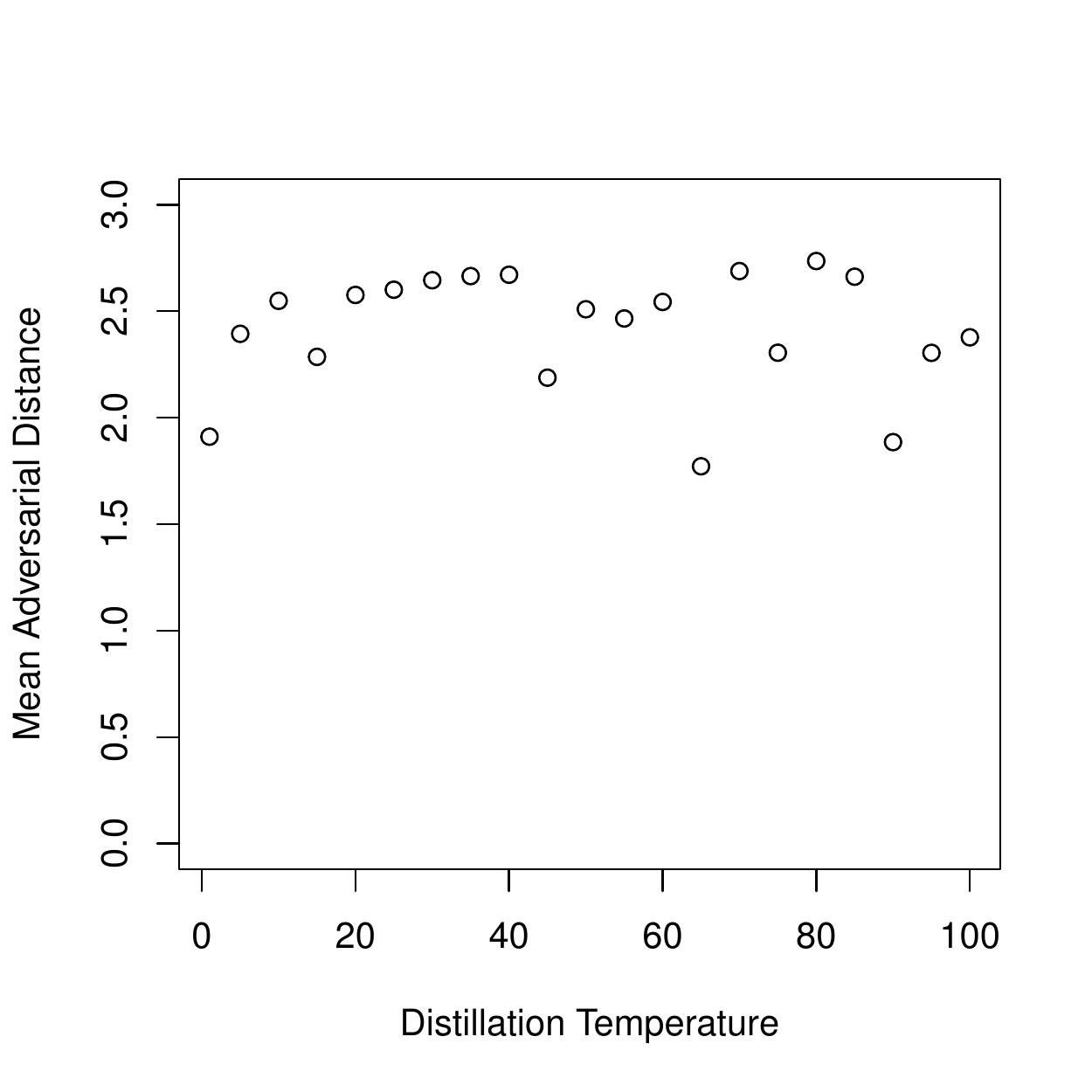}
\caption{Mean distance to targeted (with random target) adversarial examples
  for different distillation temperatures on MNIST. Temperature is uncorrelated with
  mean adversarial example distance.}
  \label{tbl:temperature}
\end{figure}

\begin{figure}[t]
  \centering
  \includegraphics[scale=.66667]{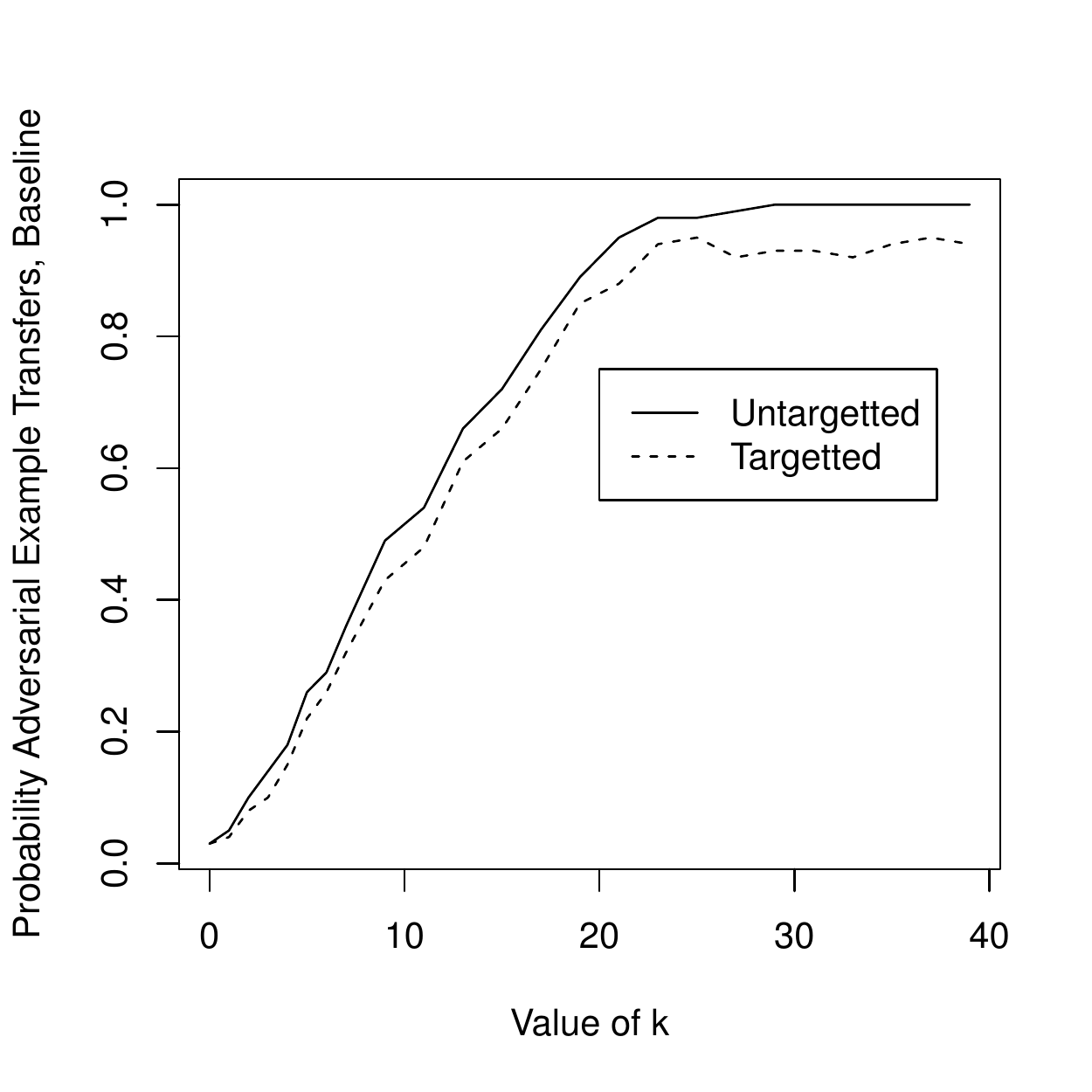}
  \caption{Probability that adversarial examples transfer from one model to another,
    for both targeted (the adversarial class remains the same) and untargeted (the
    image is not the correct class).}
  \label{tbl:transfer}
\end{figure}

\begin{figure}[t]
  \centering
  \includegraphics[scale=.66667]{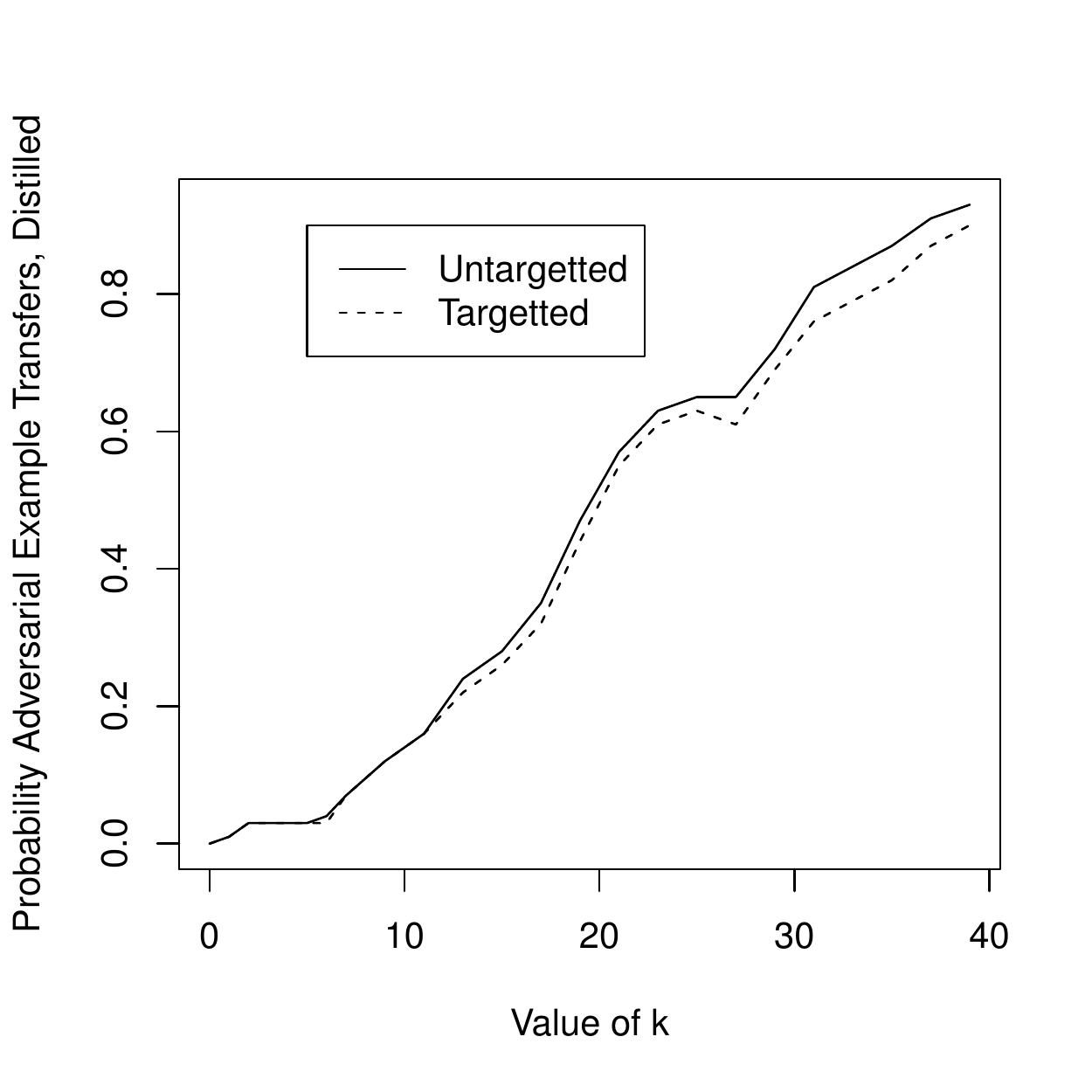}
  \caption{Probability that adversarial examples transfer from the baseline model to
  a model trained with defensive distillation at temperature 100.}
  \label{tbl:transferdistil}
\end{figure}

\subsection{Transferability}
\label{sec:transfer}

Recent work has shown that an adversarial example for one model will often
\emph{transfer} to be an adversarial on a different model, even if they are trained
on different sets of training data
\cite{szegedy2013intriguing,goodfellow2014explaining},
and even if they use entirely different
algorithms (i.e., adversarial examples on neural networks transfer to random forests \cite{papernot2016transferability}).

Therefore, any defense that is able to provide robustness against adversarial
examples \emph{must} somehow break this transferability property; otherwise, we
could run our attack algorithm on an easy-to-attack model, and then transfer those
adversarial examples to the hard-to-attack model.

Even though defensive distillation is not robust to our stronger attacks, we demonstrate a
second break of distillation by transferring attacks from a standard model to a
defensively distilled model.

We accomplish this by finding \emph{high-confidence adversarial examples}, which we define
as adversarial examples that are
strongly misclassified by the original model. Instead of looking for an adversarial
example that just barely
changes the classification from the source to the target, we want one where the target
is much more likely than any other label.

Recall the loss function defined earlier for $L_2$ attacks:
\begin{equation*}
f(x') = \max(\max \{ Z(x')_i : i \ne t\} - Z(x')_t, -\kappa).
\end{equation*}
The purpose of the parameter $\kappa$ is to control the strength
of adversarial examples: the larger $\kappa$, the stronger the classification of the
adversarial example.
This allows us to generate high-confidence adversarial examples by
increasing $\kappa$.

We first investigate if our hypothesis is true that the stronger the classification on the
first model, the more likely it will transfer. We do this by varying $\kappa$
from $0$ to $40$.

Our baseline experiment uses two models trained on MNIST as described
in Section~\ref{sec:model}, with each model trained on half of the training data.
We find that the transferability success rate increases linearly from
$\kappa=0$ to $\kappa=20$ and then plateaus at near-$100\%$ success for $\kappa \approx 20$, so clearly increasing
$\kappa$ increases the probability of a successful transferable attack.

We then run this same experiment only instead we train the second model with
defensive distillation, and find that adversarial examples \emph{do} transfer. This gives
us another attack technique for finding adversarial examples on distilled networks.

However, interestingly, the transferability success rate between the unsecured model and
the distilled model only reaches $100\%$ success at $\kappa=40$, in comparison to the
previous approach that only required $\kappa=20$.

We believe that this approach can be used in general to evaluate the robustness of
defenses, even if the defense is able to completely block flow of gradients to
cause our gradient-descent based approaches from succeeding.

\section{Conclusion}

The existence of adversarial examples limits the areas in which deep learning can be
applied. It is an open problem to construct defenses that are robust to adversarial
examples. In an attempt to solve this problem, defensive distillation was proposed
as a general-purpose procedure to increase the robustness of an arbitrary neural
network.

In this paper, we propose powerful attacks that defeat defensive distillation,
demonstrating that our attacks more generally can be used to evaluate the efficacy
of potential defenses.
By systematically evaluating many possible attack approaches, we settle on one that
can consistently find better adversarial examples than all existing approaches.
We use this evaluation as the basis of our three $L_0$, $L_2$, and $L_\infty$ attacks.

We encourage those who create defenses to perform the two evaluation approaches
we use in this paper:
\begin{itemize}
\item \textbf{Use a powerful attack} (such as the ones proposed in this paper) to evaluate the robustness of the secured model
  directly. Since a defense that prevents our $L_2$ attack will prevent our other attacks,
  defenders should make sure to establish robustness against the $L_2$ distance metric.
\item \textbf{Demonstrate that transferability fails} by constructing high-confidence
  adversarial examples on a unsecured model and showing they fail to transfer to
  the secured model.
\end{itemize}

\section*{Acknowledgements}

We would like to thank Nicolas Papernot discussing our defensive distillation
implementation, and the anonymous reviewers for their helpful feedback.
This work was supported by Intel through the ISTC for Secure Computing,
Qualcomm, Cisco, the AFOSR under MURI award FA9550-12-1-0040, and
the Hewlett Foundation through the Center for Long-Term Cybersecurity.

{\footnotesize
\bibliographystyle{acm}
\bibliography{paper}

\begin{thebibliography}{10}

\bibitem{andor2016globally}
{\sc Andor, D., Alberti, C., Weiss, D., Severyn, A., Presta, A., Ganchev, K.,
  Petrov, S., and Collins, M.}
\newblock Globally normalized transition-based neural networks.
\newblock {\em arXiv preprint arXiv:1603.06042\/} (2016).

\bibitem{bastani2016measuring}
{\sc Bastani, O., Ioannou, Y., Lampropoulos, L., Vytiniotis, D., Nori, A., and
  Criminisi, A.}
\newblock Measuring neural net robustness with constraints.
\newblock {\em arXiv preprint arXiv:1605.07262\/} (2016).

\bibitem{bojarski2016end}
{\sc Bojarski, M., Del~Testa, D., Dworakowski, D., Firner, B., Flepp, B.,
  Goyal, P., Jackel, L.~D., Monfort, M., Muller, U., Zhang, J., et~al.}
\newblock End to end learning for self-driving cars.
\newblock {\em arXiv preprint arXiv:1604.07316\/} (2016).

\bibitem{selfdriving}
{\sc Bourzac, K.}
\newblock Bringing big neural networks to self-driving cars, smartphones, and
  drones.
\newblock
  \url{http://spectrum.ieee.org/computing/embedded-systems/bringing-big-neural-networks-to-selfdriving-cars-smartphones-and-drones},
  2016.

\bibitem{carlini2016hidden}
{\sc Carlini, N., Mishra, P., Vaidya, T., Zhang, Y., Sherr, M., Shields, C.,
  Wagner, D., and Zhou, W.}
\newblock Hidden voice commands.
\newblock In {\em 25th USENIX Security Symposium (USENIX Security 16), Austin,
  TX\/} (2016).

\bibitem{chandola2009anomaly}
{\sc Chandola, V., Banerjee, A., and Kumar, V.}
\newblock Anomaly detection: A survey.
\newblock {\em ACM computing surveys (CSUR) 41}, 3 (2009), 15.

\bibitem{clevert2015fast}
{\sc Clevert, D.-A., Unterthiner, T., and Hochreiter, S.}
\newblock Fast and accurate deep network learning by exponential linear units
  ({ELUs}).
\newblock {\em arXiv preprint arXiv:1511.07289\/} (2015).

\bibitem{dahl2013large}
{\sc Dahl, G.~E., Stokes, J.~W., Deng, L., and Yu, D.}
\newblock Large-scale malware classification using random projections and
  neural networks.
\newblock In {\em 2013 IEEE International Conference on Acoustics, Speech and
  Signal Processing\/} (2013), IEEE, pp.~3422--3426.

\bibitem{deng2009imagenet}
{\sc Deng, J., Dong, W., Socher, R., Li, L.-J., Li, K., and Fei-Fei, L.}
\newblock Imagenet: A large-scale hierarchical image database.
\newblock In {\em Computer Vision and Pattern Recognition, 2009. CVPR 2009.
  IEEE Conference on\/} (2009), IEEE, pp.~248--255.

\bibitem{giusti2016machine}
{\sc Giusti, A., Guzzi, J., Cire{\c{s}}an, D.~C., He, F.-L., Rodr{\'\i}guez,
  J.~P., Fontana, F., Faessler, M., Forster, C., Schmidhuber, J., Di~Caro, G.,
  et~al.}
\newblock A machine learning approach to visual perception of forest trails for
  mobile robots.
\newblock {\em IEEE Robotics and Automation Letters 1}, 2 (2016), 661--667.

\bibitem{goodfellow2014explaining}
{\sc Goodfellow, I.~J., Shlens, J., and Szegedy, C.}
\newblock Explaining and harnessing adversarial examples.
\newblock {\em arXiv preprint arXiv:1412.6572\/} (2014).

\bibitem{graham2014fractional}
{\sc Graham, B.}
\newblock Fractional max-pooling.
\newblock {\em arXiv preprint arXiv:1412.6071\/} (2014).

\bibitem{graves2013speech}
{\sc Graves, A., Mohamed, A.-r., and Hinton, G.}
\newblock Speech recognition with deep recurrent neural networks.
\newblock In {\em 2013 IEEE international conference on acoustics, speech and
  signal processing\/} (2013), IEEE, pp.~6645--6649.

\bibitem{grosse2016adversarial}
{\sc Grosse, K., Papernot, N., Manoharan, P., Backes, M., and McDaniel, P.}
\newblock Adversarial perturbations against deep neural networks for malware
  classification.
\newblock {\em arXiv preprint arXiv:1606.04435\/} (2016).

\bibitem{gu2014towards}
{\sc Gu, S., and Rigazio, L.}
\newblock Towards deep neural network architectures robust to adversarial
  examples.
\newblock {\em arXiv preprint arXiv:1412.5068\/} (2014).

\bibitem{he2016deep}
{\sc He, K., Zhang, X., Ren, S., and Sun, J.}
\newblock Deep residual learning for image recognition.
\newblock In {\em Proceedings of the IEEE Conference on Computer Vision and
  Pattern Recognition\/} (2016), pp.~770--778.

\bibitem{38131}
{\sc Hinton, G., Deng, L., Yu, D., Dahl, G., rahman Mohamed, A., Jaitly, N.,
  Senior, A., Vanhoucke, V., Nguyen, P., Sainath, T., and Kingsbury, B.}
\newblock Deep neural networks for acoustic modeling in speech recognition.
\newblock {\em Signal Processing Magazine\/} (2012).

\bibitem{hinton2012deep}
{\sc Hinton, G., Deng, L., Yu, D., Dahl, G.~E., Mohamed, A.-r., Jaitly, N.,
  Senior, A., Vanhoucke, V., Nguyen, P., Sainath, T.~N., et~al.}
\newblock Deep neural networks for acoustic modeling in speech recognition: The
  shared views of four research groups.
\newblock {\em IEEE Signal Processing Magazine 29}, 6 (2012), 82--97.

\bibitem{hinton2015distilling}
{\sc Hinton, G., Vinyals, O., and Dean, J.}
\newblock Distilling the knowledge in a neural network.
\newblock {\em arXiv preprint arXiv:1503.02531\/} (2015).

\bibitem{huang2015learning}
{\sc Huang, R., Xu, B., Schuurmans, D., and Szepesv{\'a}ri, C.}
\newblock Learning with a strong adversary.
\newblock {\em CoRR, abs/1511.03034\/} (2015).

\bibitem{huang2016safety}
{\sc Huang, X., Kwiatkowska, M., Wang, S., and Wu, M.}
\newblock Safety verification of deep neural networks.
\newblock {\em arXiv preprint arXiv:1610.06940\/} (2016).

\bibitem{janglova2005neural}
{\sc Janglov{\'a}, D.}
\newblock Neural networks in mobile robot motion.
\newblock {\em Cutting Edge Robotics 1}, 1 (2005), 243.

\bibitem{kingma2014adam}
{\sc Kingma, D., and Ba, J.}
\newblock Adam: A method for stochastic optimization.
\newblock {\em arXiv preprint arXiv:1412.6980\/} (2014).

\bibitem{krizhevsky2009learning}
{\sc Krizhevsky, A., and Hinton, G.}
\newblock Learning multiple layers of features from tiny images.

\bibitem{krizhevsky2012imagenet}
{\sc Krizhevsky, A., Sutskever, I., and Hinton, G.~E.}
\newblock {ImageNet} classification with deep convolutional neural networks.
\newblock In {\em Advances in neural information processing systems\/} (2012),
  pp.~1097--1105.

\bibitem{kurakin2016adversarial}
{\sc Kurakin, A., Goodfellow, I., and Bengio, S.}
\newblock Adversarial examples in the physical world.
\newblock {\em arXiv preprint arXiv:1607.02533\/} (2016).

\bibitem{lecun1998gradient}
{\sc LeCun, Y., Bottou, L., Bengio, Y., and Haffner, P.}
\newblock Gradient-based learning applied to document recognition.
\newblock {\em Proceedings of the IEEE 86}, 11 (1998), 2278--2324.

\bibitem{lecun1998mnist}
{\sc LeCun, Y., Cortes, C., and Burges, C.~J.}
\newblock The mnist database of handwritten digits, 1998.

\bibitem{maas2013rectifier}
{\sc Maas, A.~L., Hannun, A.~Y., and Ng, A.~Y.}
\newblock Rectifier nonlinearities improve neural network acoustic models.
\newblock In {\em Proc. ICML\/} (2013), vol.~30.

\bibitem{melicher2016fast}
{\sc Melicher, W., Ur, B., Segreti, S.~M., Komanduri, S., Bauer, L., Christin,
  N., and Cranor, L.~F.}
\newblock Fast, lean and accurate: Modeling password guessability using neural
  networks.
\newblock In {\em Proceedings of USENIX Security\/} (2016).

\bibitem{mishkin2015all}
{\sc Mishkin, D., and Matas, J.}
\newblock All you need is a good init.
\newblock {\em arXiv preprint arXiv:1511.06422\/} (2015).

\bibitem{mnih2013playing}
{\sc Mnih, V., Kavukcuoglu, K., Silver, D., Graves, A., Antonoglou, I.,
  Wierstra, D., and Riedmiller, M.}
\newblock Playing {Atari} with deep reinforcement learning.
\newblock {\em arXiv preprint arXiv:1312.5602\/} (2013).

\bibitem{mnih2015human}
{\sc Mnih, V., Kavukcuoglu, K., Silver, D., Rusu, A.~A., Veness, J., Bellemare,
  M.~G., Graves, A., Riedmiller, M., Fidjeland, A.~K., Ostrovski, G., et~al.}
\newblock Human-level control through deep reinforcement learning.
\newblock {\em Nature 518}, 7540 (2015), 529--533.

\bibitem{moosavi2015deepfool}
{\sc Moosavi-Dezfooli, S.-M., Fawzi, A., and Frossard, P.}
\newblock Deepfool: a simple and accurate method to fool deep neural networks.
\newblock {\em arXiv preprint arXiv:1511.04599\/} (2015).

\bibitem{papernot2016cleverhans}
{\sc Papernot, N., Goodfellow, I., Sheatsley, R., Feinman, R., and McDaniel,
  P.}
\newblock cleverhans v1.0.0: an adversarial machine learning library.
\newblock {\em arXiv preprint arXiv:1610.00768\/} (2016).

\bibitem{papernot2016effectiveness}
{\sc Papernot, N., and McDaniel, P.}
\newblock On the effectiveness of defensive distillation.
\newblock {\em arXiv preprint arXiv:1607.05113\/} (2016).

\bibitem{papernot2016transferability}
{\sc Papernot, N., McDaniel, P., and Goodfellow, I.}
\newblock Transferability in machine learning: from phenomena to black-box
  attacks using adversarial samples.
\newblock {\em arXiv preprint arXiv:1605.07277\/} (2016).

\bibitem{papernot2016limitations}
{\sc Papernot, N., McDaniel, P., Jha, S., Fredrikson, M., Celik, Z.~B., and
  Swami, A.}
\newblock The limitations of deep learning in adversarial settings.
\newblock In {\em 2016 IEEE European Symposium on Security and Privacy
  (EuroS\&P)\/} (2016), IEEE, pp.~372--387.

\bibitem{distillation}
{\sc Papernot, N., McDaniel, P., Wu, X., Jha, S., and Swami, A.}
\newblock Distillation as a defense to adversarial perturbations against deep
  neural networks.
\newblock {\em IEEE Symposium on Security and Privacy\/} (2016).

\bibitem{pascanu2015malware}
{\sc Pascanu, R., Stokes, J.~W., Sanossian, H., Marinescu, M., and Thomas, A.}
\newblock Malware classification with recurrent networks.
\newblock In {\em 2015 IEEE International Conference on Acoustics, Speech and
  Signal Processing (ICASSP)\/} (2015), IEEE, pp.~1916--1920.

\bibitem{imagenet}
{\sc Russakovsky, O., Deng, J., Su, H., Krause, J., Satheesh, S., Ma, S.,
  Huang, Z., Karpathy, A., Khosla, A., Bernstein, M., Berg, A.~C., and Fei-Fei,
  L.}
\newblock {ImageNet Large Scale Visual Recognition Challenge}.
\newblock {\em International Journal of Computer Vision (IJCV) 115}, 3 (2015),
  211--252.

\bibitem{shaham2015understanding}
{\sc Shaham, U., Yamada, Y., and Negahban, S.}
\newblock Understanding adversarial training: Increasing local stability of
  neural nets through robust optimization.
\newblock {\em arXiv preprint arXiv:1511.05432\/} (2015).

\bibitem{silver2016mastering}
{\sc Silver, D., Huang, A., Maddison, C.~J., Guez, A., Sifre, L., Van
  Den~Driessche, G., Schrittwieser, J., Antonoglou, I., Panneershelvam, V.,
  Lanctot, M., et~al.}
\newblock Mastering the game of {Go} with deep neural networks and tree search.
\newblock {\em Nature 529}, 7587 (2016), 484--489.

\bibitem{springenberg2014striving}
{\sc Springenberg, J.~T., Dosovitskiy, A., Brox, T., and Riedmiller, M.}
\newblock Striving for simplicity: The all convolutional net.
\newblock {\em arXiv preprint arXiv:1412.6806\/} (2014).

\bibitem{szegedy2015rethinking}
{\sc Szegedy, C., Vanhoucke, V., Ioffe, S., Shlens, J., and Wojna, Z.}
\newblock Rethinking the {Inception} architecture for computer vision.
\newblock {\em arXiv preprint arXiv:1512.00567\/} (2015).

\bibitem{szegedy2013intriguing}
{\sc Szegedy, C., Zaremba, W., Sutskever, I., Bruna, J., Erhan, D., Goodfellow,
  I., and Fergus, R.}
\newblock Intriguing properties of neural networks.
\newblock {\em ICLR\/} (2013).

\bibitem{warde2016adversarial}
{\sc Warde-Farley, D., and Goodfellow, I.}
\newblock Adversarial perturbations of deep neural networks.
\newblock {\em Advanced Structured Prediction, T. Hazan, G. Papandreou, and D.
  Tarlow, Eds\/} (2016).

\bibitem{yuan2014droid}
{\sc Yuan, Z., Lu, Y., Wang, Z., and Xue, Y.}
\newblock Droid-sec: Deep learning in android malware detection.
\newblock In {\em ACM SIGCOMM Computer Communication Review\/} (2014), vol.~44,
  ACM, pp.~371--372.

\end{thebibliography}
}

\appendix

\section{CIFAR-10 Source-Target Attacks}

\begin{figure*}
  \hspace{1cm}
      \begin{tabular}{llllllllll}
        \multicolumn{10}{c}{Target Classification ($L_0$)} \\
        0\,\,\,\,\,\,\,\,\,\,\,\,\,\,\,\,\,\, & 1\,\,\,\,\,\,\,\,\,\,\,\,\,\,\,\,\,\, & 2\,\,\,\,\,\,\,\,\,\,\,\,\,\,\,\,\,\, & 3\,\,\,\,\,\,\,\,\,\,\,\,\,\,\,\,\,\, & 4\,\,\,\,\,\,\,\,\,\,\,\,\,\,\,\,\,\,\, & 5\,\,\,\,\,\,\,\,\,\,\,\,\,\,\,\,\,\, & 6\,\,\,\,\,\,\,\,\,\,\,\,\,\,\,\,\,\, & 7\,\,\,\,\,\,\,\,\,\,\,\,\,\,\,\,\,\, & 8\,\,\,\,\,\,\,\,\,\,\,\,\,\,\,\,\,\, & 9\, \\
  \end{tabular}  \\
  {\rotatebox[origin=l]{90}{
      \begin{tabular}{llllllllll}
        \multicolumn{10}{c}{Source Classification} \\
        \,\,\,\,\,\,\,\,\,\,\,\,0\,\,\,\,\,\,\,\,\,\,\,\,\,\,\,\,\,\, & 1\,\,\,\,\,\,\,\,\,\,\,\,\,\,\,\,\,\, & 2\,\,\,\,\,\,\,\,\,\,\,\,\,\,\,\,\,\, & 3\,\,\,\,\,\,\,\,\,\,\,\,\,\,\,\,\,\, & 4\,\,\,\,\,\,\,\,\,\,\,\,\,\,\,\,\,\,\, & 5\,\,\,\,\,\,\,\,\,\,\,\,\,\,\,\,\,\, & 6\,\,\,\,\,\,\,\,\,\,\,\,\,\,\,\,\,\, & 7\,\,\,\,\,\,\,\,\,\,\,\,\,\,\,\,\,\, & 8\,\,\,\,\,\,\,\,\,\,\,\,\,\,\,\,\,\, & 9\, \\
  \end{tabular}}}
  \centering
  \includegraphics[scale=0.35]{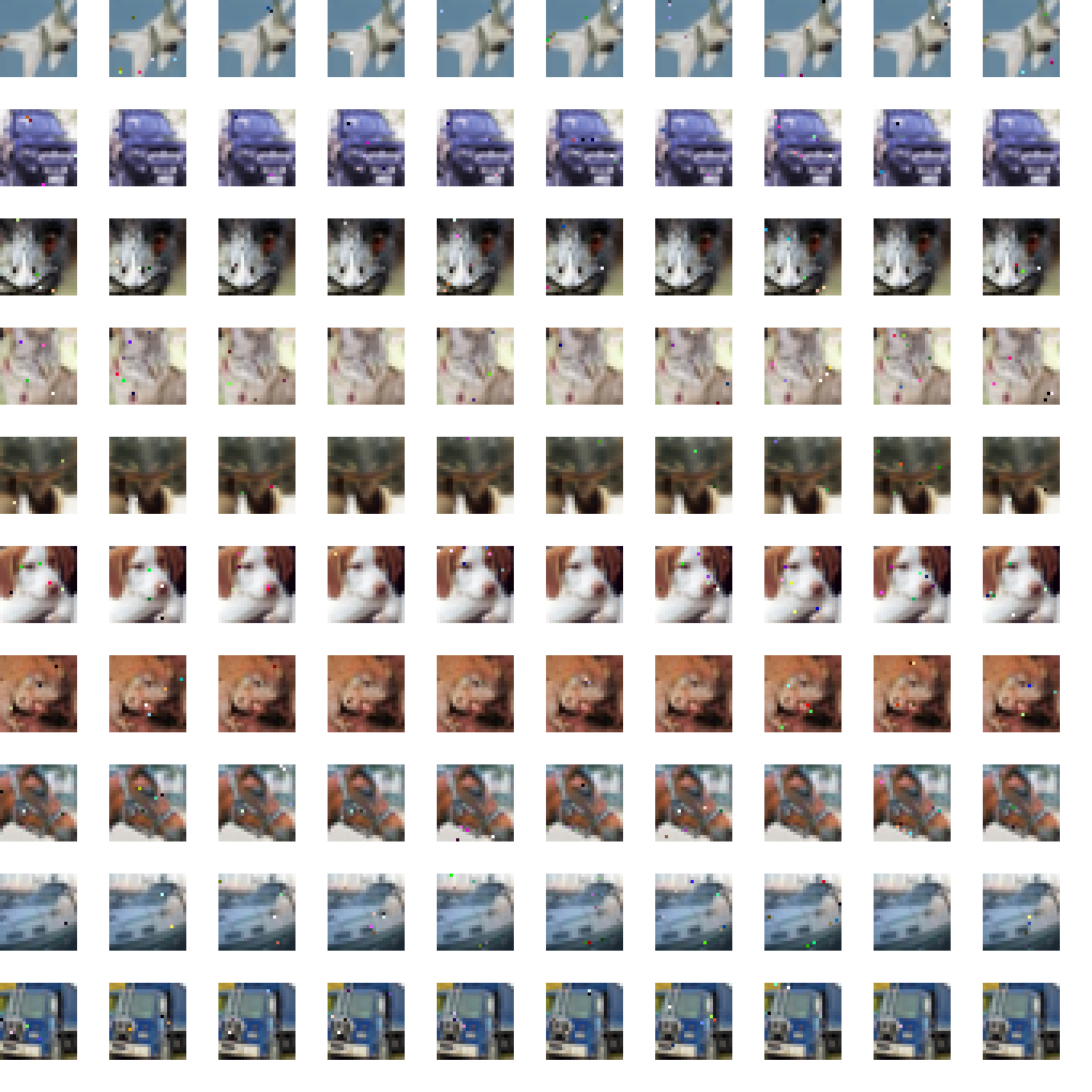}
  \caption{Our $L_0$ adversary applied to the CIFAR dataset performing a targeted attack
    for every source/target pair. Each image is the first image in the dataset with
    that label.}
  \label{fig:cifarl0}
\end{figure*}

\begin{figure*}
  \hspace{1cm}
      \begin{tabular}{llllllllll}
        \multicolumn{10}{c}{Target Classification ($L_2$)} \\
        0\,\,\,\,\,\,\,\,\,\,\,\,\,\,\,\,\,\, & 1\,\,\,\,\,\,\,\,\,\,\,\,\,\,\,\,\,\, & 2\,\,\,\,\,\,\,\,\,\,\,\,\,\,\,\,\,\, & 3\,\,\,\,\,\,\,\,\,\,\,\,\,\,\,\,\,\, & 4\,\,\,\,\,\,\,\,\,\,\,\,\,\,\,\,\,\,\, & 5\,\,\,\,\,\,\,\,\,\,\,\,\,\,\,\,\,\, & 6\,\,\,\,\,\,\,\,\,\,\,\,\,\,\,\,\,\, & 7\,\,\,\,\,\,\,\,\,\,\,\,\,\,\,\,\,\, & 8\,\,\,\,\,\,\,\,\,\,\,\,\,\,\,\,\,\, & 9\, \\
  \end{tabular}  \\
  {\rotatebox[origin=l]{90}{
      \begin{tabular}{llllllllll}
        \multicolumn{10}{c}{Source Classification} \\
        \,\,\,\,\,\,\,\,\,\,\,\,0\,\,\,\,\,\,\,\,\,\,\,\,\,\,\,\,\,\, & 1\,\,\,\,\,\,\,\,\,\,\,\,\,\,\,\,\,\, & 2\,\,\,\,\,\,\,\,\,\,\,\,\,\,\,\,\,\, & 3\,\,\,\,\,\,\,\,\,\,\,\,\,\,\,\,\,\, & 4\,\,\,\,\,\,\,\,\,\,\,\,\,\,\,\,\,\,\, & 5\,\,\,\,\,\,\,\,\,\,\,\,\,\,\,\,\,\, & 6\,\,\,\,\,\,\,\,\,\,\,\,\,\,\,\,\,\, & 7\,\,\,\,\,\,\,\,\,\,\,\,\,\,\,\,\,\, & 8\,\,\,\,\,\,\,\,\,\,\,\,\,\,\,\,\,\, & 9\, \\
  \end{tabular}}}
  \centering
  \includegraphics[scale=0.35]{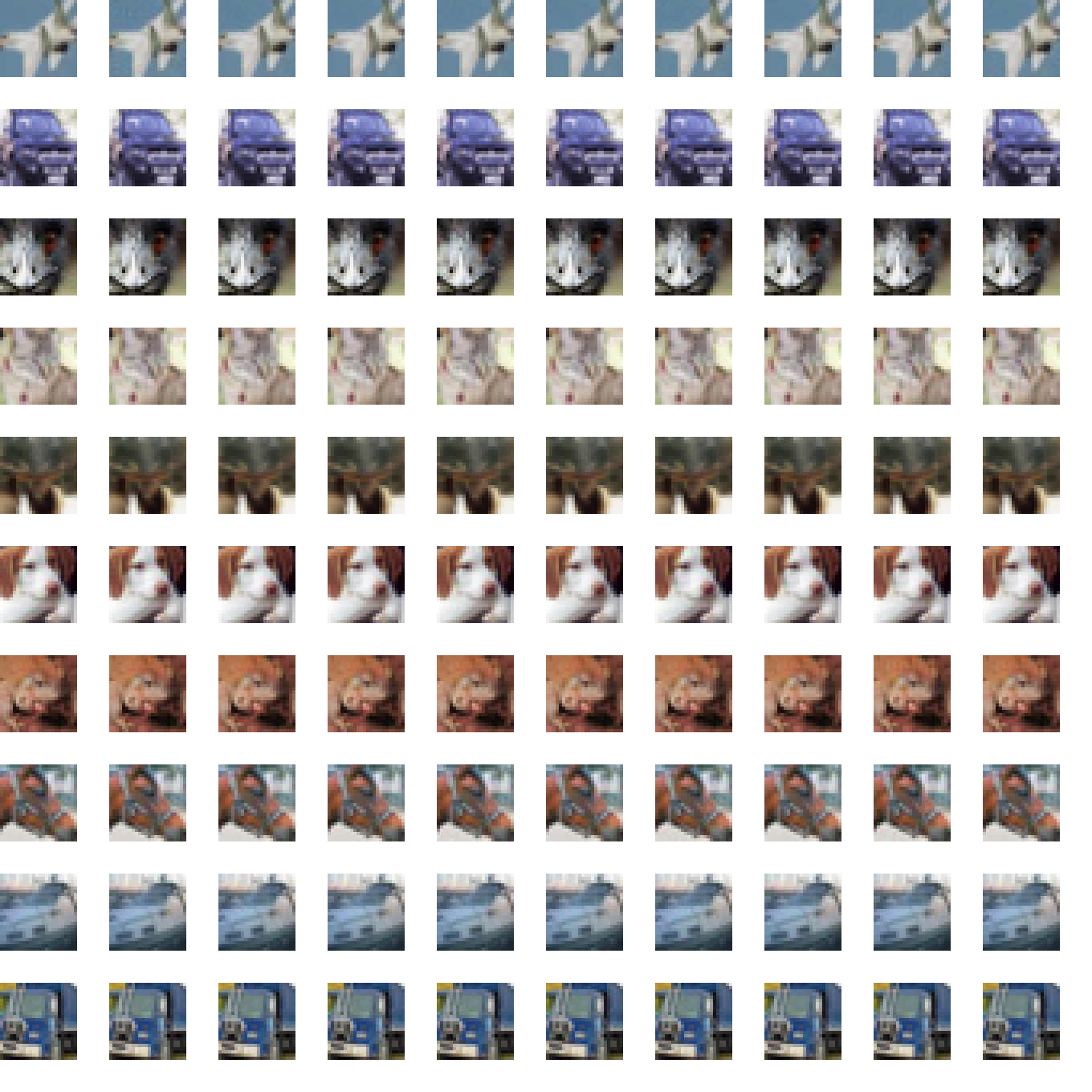}
  \caption{Our $L_2$ adversary applied to the CIFAR dataset performing a targeted attack
    for every source/target pair. Each image is the first image in the dataset with
    that label.}
  \label{fig:cifarl2}
\end{figure*}

\begin{figure*}
  \hspace{1cm}
      \begin{tabular}{llllllllll}
        \multicolumn{10}{c}{Target Classification ($L_\infty$)} \\
        0\,\,\,\,\,\,\,\,\,\,\,\,\,\,\,\,\,\, & 1\,\,\,\,\,\,\,\,\,\,\,\,\,\,\,\,\,\, & 2\,\,\,\,\,\,\,\,\,\,\,\,\,\,\,\,\,\, & 3\,\,\,\,\,\,\,\,\,\,\,\,\,\,\,\,\,\, & 4\,\,\,\,\,\,\,\,\,\,\,\,\,\,\,\,\,\,\, & 5\,\,\,\,\,\,\,\,\,\,\,\,\,\,\,\,\,\, & 6\,\,\,\,\,\,\,\,\,\,\,\,\,\,\,\,\,\, & 7\,\,\,\,\,\,\,\,\,\,\,\,\,\,\,\,\,\, & 8\,\,\,\,\,\,\,\,\,\,\,\,\,\,\,\,\,\, & 9\, \\
  \end{tabular}  \\
  {\rotatebox[origin=l]{90}{
      \begin{tabular}{llllllllll}
        \multicolumn{10}{c}{Source Classification} \\
        \,\,\,\,\,\,\,\,\,\,\,\,0\,\,\,\,\,\,\,\,\,\,\,\,\,\,\,\,\,\, & 1\,\,\,\,\,\,\,\,\,\,\,\,\,\,\,\,\,\, & 2\,\,\,\,\,\,\,\,\,\,\,\,\,\,\,\,\,\, & 3\,\,\,\,\,\,\,\,\,\,\,\,\,\,\,\,\,\, & 4\,\,\,\,\,\,\,\,\,\,\,\,\,\,\,\,\,\,\, & 5\,\,\,\,\,\,\,\,\,\,\,\,\,\,\,\,\,\, & 6\,\,\,\,\,\,\,\,\,\,\,\,\,\,\,\,\,\, & 7\,\,\,\,\,\,\,\,\,\,\,\,\,\,\,\,\,\, & 8\,\,\,\,\,\,\,\,\,\,\,\,\,\,\,\,\,\, & 9\, \\
  \end{tabular}}}
  \centering
  \includegraphics[scale=0.35]{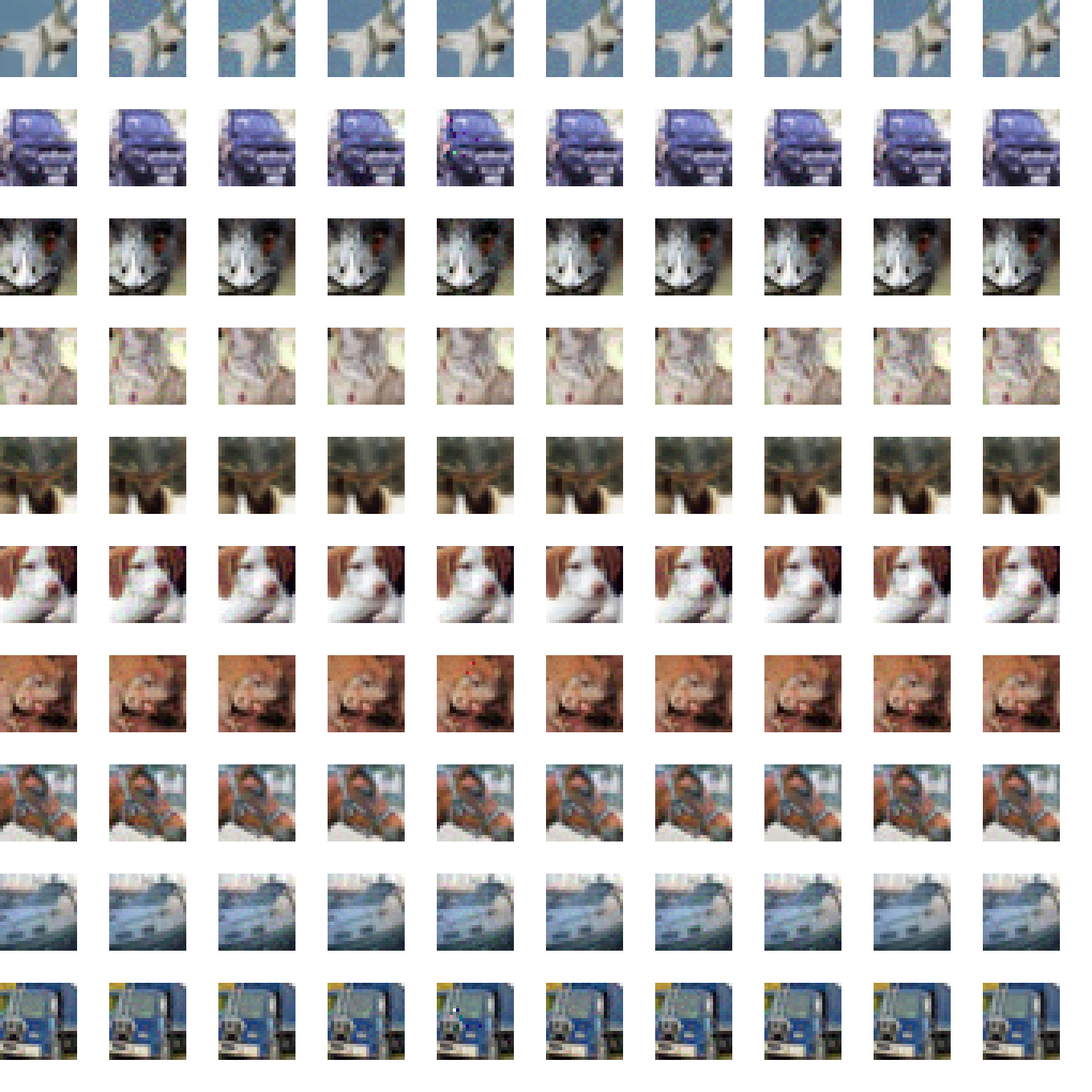}
  \caption{Our $L_\infty$ adversary applied to the CIFAR dataset performing a targeted attack
    for every source/target pair. Each image is the first image in the dataset with
    that label.}
  \label{fig:cifarli}
\end{figure*}
\end{document}